\documentclass[iop,apj]{emulateapj}
\usepackage{times}
\usepackage{textcomp}
\usepackage[normalem]{ulem}
\usepackage{amsmath,amssymb,amstext}
\usepackage[breaklinks,colorlinks,citecolor=blue,linkcolor=magenta]{hyperref} 
\usepackage[all]{hypcap} 

\shorttitle{The ELQS North Galactic Cap Sample}
 
\shortauthors{Schindler et al.}

\begin{document}

\title{The Extremely Luminous Quasar Survey (ELQS) in the Sloan Digital Sky Survey footprint. II. The North Galactic Cap Sample }

\author{Jan-Torge Schindler\altaffilmark{1}}
\author{Xiaohui Fan\altaffilmark{1}}
\author{Ian D. McGreer\altaffilmark{1}}
\author{Jinyi Yang\altaffilmark{1}}
\author{Feige Wang\altaffilmark{2}}
\author{Richard Green\altaffilmark{1}}

\author{Nicolas Garavito-Camargo\altaffilmark{1}}
\author{Yun-Hsin Huang\altaffilmark{1}}
\author{Christine O'Donnell\altaffilmark{1}}
\author{Anna Patej\altaffilmark{1}}
\author{Ragadeepika Pucha\altaffilmark{1}}
\author{Jon M. Rees\altaffilmark{1}}
\author{Eckhart Spalding\altaffilmark{1}}

\altaffiltext{1}{Steward Observatory, University of Arizona, 933 North Cherry Avenue, Tucson, AZ 85721,USA}
\altaffiltext{2}{Department of Physics, University of California Santa Barbara,  Santa Barbara, CA 93106-9530, USA}



\defcitealias{Schindler2017}{Paper I}

\begin{abstract}

We present the North Galactic Cap sample of the Extremely Luminous Quasar Survey (ELQS-N), which targets quasars with $M_{1450}<-27$ at $2.8 \leq z < 5$ in an area of $\sim7600\,\rm{deg}^2$ of the Sloan Digital Sky Survey (SDSS) footprint with $90\text{\textdegree}<\rm{RA}<270\text{\textdegree}$.
Based on a near-infrared/infrared \textit{JKW2} color cut, the ELQS selection efficiently uses random forest methods to classify quasars and to estimate photometric redshifts;  this scheme overcomes some of the difficulties of pure optical quasar selection at $z\approx3$. 
As a result, we retain a completeness of $>70\%$ over $z\sim3.0-5.0$ at $m_{i}\lesssim17.5$, limited toward fainter magnitudes by the depth of the Two Micron All Sky Survey (2MASS).
The presented quasar catalog consists of a total of 270 objects, of which 39 are newly identified in this work with spectroscopy obtained at the Vatican Advanced Technology Telescope and the MMT $6.5\,\rm{m}$ telescope.
In addition to the high completeness, which allowed us to discover new quasars in the already well-surveyed SDSS North Galactic Cap, the efficiency of our selection is relatively high at $\sim79\%$.
Using 120 objects of this quasar sample we are able to extend the previously measured optical quasar luminosity function (QLF) by one magnitude toward the bright end at $2.8 \leq z \leq 4.5$. A first analysis of the QLF suggests a relatively steep bright-end slope of $\beta\approx-4$ for this sample. This result contrasts with previous results in the same redshift range, which find a much flatter slope around $\beta\sim-2.5$, but agrees with recent measurements of the bright-end slope at lower and higher redshifts. Our results constrain the bright-end slope at $z=2.8-4.5$ to $\beta<-2.94$ with a 99\% confidence.

\end{abstract}

\keywords{galaxies: nuclei - galaxies: active - galaxies: high-redshift - quasars: general} 

\section{Introduction}


Quasars, the rapidly accreting supermassive black holes (SMBHs) at the centers of galaxies, are the most luminous non-transient light sources in the Universe. They play an important role in the formation and evolution of galaxies as the mass of the SMBH seems to be closely related to the properties of the host galaxy \citep[see][for a review]{Kormendy2013}. As strong light beacons they further allow astronomers to probe the large-scale structure formation of the universe and the nature of the intergalactic medium \citep[IGM; ][]{Simcoe2004, Prochaska2005, Worseck2011}.
Quasars discovered within the first billion years of the universe \citep{Mortlock2011, Banados2018} place strong constraints on the formation and growth of SMBHs.

The Sloan Digital Sky Survey \citep[SDSS;][]{York2000}, the SDSS Baryon Oscillation Spectroscopic Survey (BOSS; \citet{Eisenstein2011, Dawson2013}) and the SDSS extended BOSS (eBOSS; \citet{Dawson2016}) have built the largest quasar sample to date with over 500,000 objects at redshifts $z<6$. At higher redshifts quasar samples have now grown to $\sim120$ quasars \citep{Banados2016, Jiang2016} allowing statistical studies at $z>6$ for the first time.

The most important statistic for studying the formation and growth of SMBHs is the quasar luminosity function (QLF). The QLF is a measure of the spatial number density of quasars as a function of absolute magnitude (or luminosity) and redshift. The large SDSS quasar samples  provide the tightest constraints on the optical QLF so far \citep{Richards2006, Shen2012, Ross2013}.

The QLF is best described by a ``broken'' double power law \citep{Boyle1988, Boyle2000, Pei1995}, defined by a bright-end and a faint-end slope, the break point between the slopes, and an overall normalization. The faint-end slope is generally flatter than the bright-end slope and all parameters are known to change with redshift.

The bright-end slope at intermediate redshifts, $z=3-5$, has been the topic of debate in the literature. While earlier studies discovered a flattening of the slope toward higher redshifts \citep{Koo1988, Schmidt1995, Fan2001b, Richards2006}, a range of more recent studies argued that it remains steep until the highest redshifts \citep{Jiang2008, Croom2009,  Willott2010a,McGreer2013, Yang2016}.

However, it is challenging to reliably constrain the evolution of the bright-end slope, because the spatial quasar density rapidly decreases toward higher redshifts and higher luminosities. For example, at redshifts $z=3-4$ the spatial number density of quasars with absolute magnitudes $M_{1450}\lesssim -27.5$ ($m_{i}[z=2]\lesssim-29$) is less than $10^{-9}\,\rm{Mpc}^{-3}\rm{mag}^{-1}$ \citep[see][]{Richards2006, Shen2012, Ross2013}.
Therefore, wide area spectroscopic surveys like SDSS are necessary to achieve a large enough sample size to measure the bright-end slope reliably. Furthermore, in order to avoid systematic biases in the measurement of the QLF a well-constrained survey completeness and selection function is crucial.

In \citet[][hereafter Paper I]{Schindler2017} we have shown that the SDSS and BOSS quasar selection missed bright $z=3-4$ quasars due to the survey's incompleteness in the South Galactic Cap ($\rm{RA} {>} 270\text{\textdegree}$ or $\rm{RA} {<} 90\text{\textdegree}$) and difficulties in the purely optical quasar selection at these redshifts, where quasar photometry is indistinguishable from stars.

In order to more reliably constrain the bright-end QLF, we have designed the ELQS using a more inclusive quasar selection based on a near-infrared/infrared color criterion using photometry from the Two Micron All Sky Survey  \citep[2MASS][]{Skrutskie2006} and the \textit{Wide-field Infrared Survey Explorer} mission (\textit{WISE}) \citet{Wright2010}. We further apply random forests to the joint infrared and optical photometry to classify the candidates and estimate their photometric redshifts. 
We briefly summarize our quasar selection method in Section\,\ref{sec_selection_lit}. For a full discussion on the quasar selection and the resulting quasar candidate catalog, please refer to \citetalias{Schindler2017}.

In this work, we present the results of the ELQS survey in the North Galactic Cap (ELQS-N; $90\text{\textdegree} {<} \rm{RA} {<} 270\text{\textdegree}$) of the SDSS footprint. The survey is limited to objects with SDSS \textit{i}-band magnitudes $m_{i}\leq18.0$. We have followed up all ELQS-N quasar candidates, resulting in a spectroscopically complete area of $\approx7,600\,\rm{deg}^2$. Based on this sample we calculate a first estimate of the bright-end QLF at $z=2.8-4.5$ (see Section\,\ref{sec_lumfun}).

We begin with a brief description of our quasar selection process and a review of quasar catalogs from the literature used in this work (Section\,\ref{sec_selection_lit}). In Section\,\ref{sec_obs} we describe the spectroscopic observations and the data reduction. 
Subsequently, we present the ELQS-N quasar catalog in Section\,\ref{sec_elqs_spring_catalog} and discuss the ELQS selection function and the survey's completeness in Section\,\ref{sec_completeness}. We continue to describe the impact of the ELQS on the luminosity function in Section\,\ref{sec_lumfun} before we summarize our findings in Section\,\ref{sec_conclusion}

All magnitudes are displayed in the AB system \citep{Oke1983} and corrected for galactic extinction \citep{Schlafly2011} unless otherwise noted.
We denote magnitudes not corrected for galactic extinction only by $x$, where $x$ refers to the wavelength band in question, as opposed to extinction corrected magnitudes $m_{\rm{x}}$.
We adopt the standard flat $\Lambda\rm{CDM}$ cosmology with $H_0=70\,\rm{km}\rm{s}^{-1}\rm{Mpc}^{-1}$, $\Omega_{\rm{m}}=0.3$ and $\Omega_\Lambda=0.7$ in general consistent with recent measurements \citep{PlanckCollaboration2016}.

\section{ELQS Candidate Selection and Quasar Identifications from the Literature}\label{sec_selection_lit}

\subsection{The ELQS Candidate Selection}

In \citetalias{Schindler2017} we developed a quasar selection method that includes near-infrared/infrared photometry to overcome the difficulties of pure optical quasar selections. We have used 2MASS \textit{J}- and \textit{K}-band photometry as well as \textit{WISE} \textit{W2}-band photometry to devise a highly inclusive color cut: $ \rm{K}-\rm{W2} \ge 1.8 - 0.848 \cdot \left(\rm{J}-\rm{K} \right)$ (Vega magnitudes). 
The ELQS is focused on the brightest quasars ($M_{1450}<-27$) at $2.8 \leq z < 5$, which are detectable with 2MASS and \textit{WISE}.
We therefore select sources from the \textit{WISE} AllWISE Catalog, which is prematched with the 2MASS Point Source Catalog (PSC), using our \textit{JKW2} color cut and then cross match them with SDSS photometry. Extended sources are rejected using a limit on the Petrosian radius (\texttt{petroRad\_i}$\le 2\farcs0$) and we require the sample to satisfy an \textit{i}-band magnitude cut of $m_{i}\leq18.5$.

In the next step we employ the random forest method \citep{Breiman2001} to calculate photometric redshifts and further classify our candidates. Photometric redshift estimation is carried out using random forest regression trained on a quasar sample built from the SDSS DR7 and DR12 quasar catalogs \citep{Schneider2010, Paris2017}. As features we use flux ratios built from all SDSS bands and the \textit{WISE} \textit{W1} and \textit{W2} bands as well as the SDSS \textit{i}-band and \textit{WISE} \textit{W1}-band magnitudes.
To achieve a higher efficiency for the quasar selection, we also employ random forests to classify our candidates in five stellar classes according to the A, F, G, K, and M spectral types and four quasar redshift classes ($0{<}z{\leq}1.5$, $1.5{<}z{\leq}2.2$, $2.2{<}z{\leq}3.5$, and $3.5{<}z$). In this case, the training set is built from the SDSS DR7 and DR12 quasar catalogs and a sample of spectroscopically classified stars based on SDSS DR13. For the classification we have used the same features as described above for the regression. We apply our quasar selection to the full SDSS footprint, excluding the galactic plane ($b<-20$ or $b>30$).

Our primary high-priority sample encompasses all candidates, which are generally classified to be quasars, have a regression redshift of $z_{\rm{reg}}\geq 2.8$ and satisfy $m_{i}\leq18.0$. We identify 594 of these objects with our quasar selection method, of which 324 were previously identified in the literature. The remaining 270 candidates lack spectroscopic identification. After a visual inspection, we discard 85 candidates due to unreliable photometry. In most cases the \textit{WISE} image in the \textit{W1} or \textit{W2} bands was blended with another source or the image showed artifacts, like bright tails, detected as a source. In a few cases similar artifacts were persistent in the 2MASS or SDSS photometry.
We have followed up on 184 primary candidates in the ELQS-N and ELQS South Galactic Cap sample (ELQS-S). Data reduction and spectroscopic identification of the ELQS-S sample still continues. The ELQS-S sample will be published along with a full analysis of the QLF in Paper III (Schindler et al. 2018, in preparation). Table\,\ref{tab_cand_sample} provides an overview of the candidate sample. For further details regarding the quasar selection, we refer the interested reader to \citetalias{Schindler2017}.

\subsection{ELQS Candidates in the Literature}

The majority of previously identified ELQS candidates were observed as part of the SDSS I/II \citep{Abazajian2009}, BOSS, and eBOSS and published in the SDSS DR7 \citep{Schneider2010} and DR14 \citep{Paris2018} quasar catalogs. Further quasar identifications come from the Million Quasar Catalog (MQC) \citep {Flesch2015} and an ongoing survey by Yang et al. (2018, in preparation).

\begin{table*}[htp]
\footnotesize
\centering
 \caption{ELQS primary candidate sample}
 \label{tab_cand_sample}
 \begin{tabular}{lccc}
 \tableline
 Primary Candidates  & Full Area & ELQS-N & ELQS-S \\
  ($m_{i}\leq18.0$ and $z_{\rm{reg}}\geq2.8$)& & ($90\text{\textdegree} {<} \rm{RA} {<} 270\text{\textdegree}$) & ($\rm{RA} {>} 270\text{\textdegree}$ or $\rm{RA} {<} 90\text{\textdegree}$)\\
 \tableline
 \tableline
  Total selected primary candidates & 594 & 375 & 219 \\
 \tableline
  Good primary candidates (excluding bad photometry) & 509 & 340 & 169 \\
 \tableline
 \tableline
  Good primary candidates in the literature & 324 & 252 & 72 \\
  Good primary candidates observed & 184 & 88 & 96 \\
  Good primary candidates to observe & 1 & 0 & 1 \\
 \tableline
  Good primary candidates in the literature at $z>2.8$ & 298 & 231 & 67 \\
  Good primary candidates observed and identified as $z>2.8$ QSOs& 108 & 39 & 69\tablenotemark{a}\\
 \tableline
 \end{tabular} 
\tablenotetext{1}{Data reduction and spectroscopic identification of the ELQS-S sample is not complete.} 
\end{table*}

\subsection{SDSS DR7}

The DR7 concluded the SDSS I/II spectroscopic survey. The quasar target selection is described in \citet{Richards2002} and the data were reduced with the standard SDSS pipeline \citep{Stoughton2002}. 
The quasar target selection flagged all objects as candidates, which are outliers of the stellar locus in \textit{griz} color space, in addition to several inclusion regions to target specific redshift ranges \citep{Richards2002}.
The DR7 quasar catalog (DR7Q) presented over 100,000 spectroscopically confirmed quasars in a $\approx 9380\,\rm{deg}^2$ region of the SDSS DR7 footprint \citep{Schneider2010}. 

In our full primary candidate sample, obeying the \textit{\textit{JKW2}} color cut with $m_{i}\leq18.0$ and regression redshifts $z_{\rm{reg}}\geq2.8$, there are 203 known quasars from the DR7Q, of which 194 are at $z\geq2.8$ and 9 below $z=2.8$. 

The full DR7Q catalog contains 265 quasars with $m_{i}\leq18.0$ at $z\geq2.8$. We lose 54 objects, because they do not have the necessary 2MASS/\textit{WISE} photometry or do not make the \textit{JKW2} color cut. We miss another 17 quasars, because our redshift estimates at $2.8\leq z\leq 3.0$ are not fully accurate. Both of these effects are reflected in our selection function described in Section\,\ref{sec_completeness}.

\subsection{BOSS and eBOSS}

The BOSS quasar survey targeted quasars and galaxies to investigate baryon acoustic oscillations (BAO). Quasars in a redshift range of  $2.2 < z < 3.5$ were primarily targeted to study BAOs using the quasars themselves and their Ly$\alpha$ forests \citep{McDonald2007, Ross2012}. The BOSS survey used the extreme deconvolution algorithm \citep[XDQSO,][]{Bovy2011} to select quasar candidates, which was optimized for the targeted redshift range. 
The BOSS spectra are obtained with a fiber-fed multi-object spectrograph \citep{Smee2013} and were reduced using a pipeline described in \citet{Bolton2012}.
The BOSS quasar campaign culminated in the publication of the DR12 quasar catalog \citep[DR12Q,][]{Paris2017} with $\approx 300,000$ quasars.
The SDSS-IV extension of the BOSS observational cosmology program is the extended Baryon Oscillation Spectroscopic Survey \citep[eBOSS,][]{Dawson2016, Blanton2017}.
For the eBOSS BAO measurements quasars in a redshift range of  $0.9 < z < 2.2$ are targeted. In BOSS and eBOSS more than 60,000 quasars were discovered at $z>2.1$. The quasar selection is described in \citep{Myers2015}. Primarily the XDQSO method and a mid-IR color cut were used to form a uniform quasar sample over  $\approx 7500\,\rm{deg}^2$ with  $g < 22$ or $r < 22$.

A second selection based on quasar variability uses multi-epoch imaging from the Palomar Transient Factory to discover quasars up to $g<22.5$ at $z>2.1$.
With the 14th data release of SDSS, the first eBOSS quasar catalog \citep[DR14Q;][]{Paris2018} was published containing more than 500,000 objects. It includes all quasars identified with the BOSS and the majority of quasars from the original DR7Q.

In our full primary candidate sample, obeying the \textit{JKW2} color cut with  $m_{i}\leq18.0$ and $z_{\rm{reg}}\geq2.8$, there are 256 known quasars from DR14Q of which 240 are at $z\geq2.8$. These 240 quasars include all but one quasar from the original SDSS DR7Q.

The DR14 quasar catalog includes a total of 341 quasars with $m_{i}\leq18.0$ at $z\geq2.8$. The majority of quasars we did not recover did not have the necessary 2MASS and \textit{WISE} photometry to evaluate the \textit{JKW2} color cut. Due to our cut on the photometric redshift estimate we lose an additional 25 quasars from DR14Q, which are between $2.8\leq z \leq 3.16$. Our selection function reflects these effects (Section\,\ref{sec_completeness}).

One of our quasar candidates had a spectrum taken as part of the Segue\,2 program and was flagged as a broad-line quasar by the automated pipeline, but it was not included in any of the SDSS quasar catalogs. We visually determined it to be a BAL quasar at $z=3.65$ and discuss its properties in Section\,\ref{sec_individ_objects}.




\subsection{MQC}

The MQC \citep{Flesch2015} is a compendium of type I and II active galactic nuclei (AGN) from the available literature. It includes over 600,000 type-I AGN and quasar candidates from the  NBCKDE \citep{Richards2009}, NBCKDE-v3 \citep{Richards2015}, XDQSO \citep{Bovy2011}, AllWISE \citep{Secrest2015} and Peters \citep{Peters2015} photometric quasar catalogs as well as from all-sky radio/X-ray associated objects, which are calculated by the author. 
With a total of $\sim 2,000,000$ objects the MQC is the largest compilation of quasars and candidates to date. However, we are only using the spectroscopically confirmed quasars. 
The MQC includes all quasars in the SDSS quasar catalogs. With regard to the MQC we will only discuss quasar candidates below that do not have successful cross matches to the SDSS DR7Q, DR12Q and DR14Q.
Our primary candidate catalog over the entire SDSS footprint, excluding the galactic plane, includes 39 MQC quasars that make the \textit{JKW2} color cut and have $m_{i}\leq18.0$ and $z_{\rm{reg}}\geq2.8$, of which 31 are at $z\geq2.8$.

The ELQS-N quasar sample includes 15 validated quasars from the MQC. The redshift references of these objects are mainly from the Half Million Quasars Catalogue  \citep[12 objects, ][]{Flesch2015}, the Large Sky Area Multi-Object Fiber Spectrograph (LAMOST) DR3 (2 objects) and SDSS DR14 (1 object). The one SDSS DR14 quasar is not included in the DR14Q quasar catalog and was targeted as part of the Segue\,2 project.

\subsection{Yang et al. Bright Spectroscopic Quasar Survey}

Yang et al. (2018, in preparation) are currently conducting a spectroscopic survey similar to this work. The goals of their survey are to find bright quasars at $z\approx2-3$ and at $z\geq4$ missed by the SDSS/BOSS/eBOSS quasar surveys and to test different quasar selection criteria for the upcoming LAMOST quasar survey. 
The survey is based on two different candidate samples. The first sample uses photometry from SDSS, the UKIRT (United Kingdom Infrared Telescope) Infrared Deep Sky Survey (UKIDSS) and the Vista Hemisphere Survey. Candidates are selected using the \textit{YK}-\textit{gz} and \textit{JK}-\textit{iY} color cuts presented in \citet{Wu2010}. The second sample is based on SDSS and \textit{WISE} photometry using the methods presented in \citet{Wu2012}. Spectroscopic observations are carried out using the Lijiang telescope ($2.4\,\rm{m}$) and the Xinglong telescope ($2.16\,\rm{m}$).

In our total primary candidate sample, obeying the \textit{JKW2} color cut with $m_{i}\leq18.0$ and $z_{\rm{reg}}\geq2.8$, there are 34  new quasars identified as part of their observational campaign, of which 33 are at $z>2.8$.

The ELQS-N includes 10 quasars spectroscopically identified as part of their survey and another 8 quasars that were both identified by this work as well as their survey.

\subsection{Hubble Space Telescope (HST) GO Proposal 13013 - PI: Gabor Worseck}
In addition to the references above, all quasar candidates were matched to the NASA/IPAC Extragalactic Database (NED) to exclude other already known quasars. However, even though one of our quasar candidates, J163056.335+043559.42, was not flagged as a quasar, we later discovered that it was part of the HST GO program 13013\footnote{\url{http://www.stsci.edu/hst/phase2-public/13013.pro}} (PI: Gabor Worseck) and subsequently studied by \citet{Zheng2015}.
Although this quasar can be found in the literature, we decided to include it in our new discoveries to formally publish its classification.

\section{Spectroscopic Observations and Data Reduction}\label{sec_obs}

The 88 ELQS-N primary candidates were observed with the Vatican Advanced Technology Telescope (VATT) and the MMT $6.5\,\rm{m}$ Telescope. In this section we will provide details regarding the spectroscopic observations and the data reduction.

\subsection{VATT Observations}
The majority of the spectroscopic identifications were carried out with the VATT using the VATTSpec spectrograph. We have used the  $300\,$g/mm grating in first order blazed at $5000\,\text{\AA}$. The spectra have a resolution of $R\sim1000$ ($1\farcs5$ slit) and a coverage of $\sim4000\,\text{\AA}$ around a central wavelength of $\sim5775\,\text{\AA}$. 

The observations for the ELQS were conducted in multiple campaigns. Pilot observations started in 2015 April 21-25, with continuing observations in October 8-12. The program continued through 2016 March 10-15, April 10-13, November 19-23 and December 18-20. In 2017 we finished the North Galactic Cap footprint with observations on April 3-6, April 17-19 and May 3-5.
About 45\% percent of the awarded VATT time was lost, mainly due to bad weather conditions.

The exposure times varied between 15 and 30 minutes depending on the weather conditions and the brightness of the candidate resulting in low signal-to-noise ratio (S/N) spectra. Quasars were easily identified by their prominent emission lines.

The data were reduced using the standard long slit reduction methods within the IRAF software package \citep{Tody1986, Tody1993}. This includes bias subtraction, flat field corrections, and sky subtractions using polynomial background fits along the slit direction. The last task was carried out using the \texttt{apall} routine. All observations since 2016 October were reduced using optimal extraction (weights=variance) and cosmic-ray reduction within the apall routine.
Wavelength calibration was carried out using an internal HgAr lamp. Flux calibration was done using standard stars observed once per night. Absolute flux calibration may not be reliable in all cases due to passing clouds.
Therefore, we have adjusted the flux levels in the spectra to reflect the measured SDSS magnitudes. In all cases we used the SDSS \textit{r}-band for this re-calibration. The spectra have not been corrected for telluric absorption features.

The candidate selection for the ELQS survey was finalized in 2016 September. Previous observations include candidates with less efficient selection criteria based on simple color cuts or with a fainter limiting magnitude of $m_{i}=18.5$. Overall we have observed 325 unique candidates with the VATT, out of which 121 were generally identified as quasars. A full catalog of all observed objects, not included in the ELQS-N quasar catalog, will be published with the completion of the ELQS survey.

\subsection{MMT Observations}

We have used the MMT Red Channel Spectrograph to carry out followup observations of our newly discovered quasars. For all observations we used the MMT 270\,g/mm and 300\,g/mm gratings blazed at 1st/$7300\text{\AA}$ and 1st/$4800\text{\AA}$, respectively. Regarding the 270\,g/mm grating we used central wavelengths of $6400$ and $7150\text{\AA}$. For the 300\,g/mm we used central wavelengths of $5000$, $5500$, $5560$, $5570$, and $6083\text{\AA}$. The 270\,g/mm grating has an approximate coverage of $3705\text{\AA}$, whereas the 300\,g/mm grating has an approximate coverage of $3313\text{\AA}$. We chose exposure times of $\sim3-15\,\rm{min}$ per spectrum, depending on the object and conditions.

Based on the seeing conditions, we have either used the $1\farcs25$ or the $1\farcs5$ slit, providing a resolution of $R\approx300-400$ with both gratings.
Observations were taken in 2015 on May 9, November 8-9 and 11, and December 1 as well as in 2017 on May 17-18. The data reduction procedure is analogous to the one used for the VATT data above. For the wavelength calibration we have used the internal HeArNe lamps and spectrophotometric standards were observed once per night. The flux calibration may not be reliable due to changing weather conditions. As for all observations we have rescaled the spectral flux to match the SDSS \textit{r}-band magnitudes. 

After the completion of the survey we noticed that the MMT Red Channel Spectrograph dim continuum lamp failed during our run on 2017 May 17-18 resulting in very low S/N flat fields for those two nights. For all spectra with the 300\,g/mm grating and a central wavelength of  $5560\text{\AA}$, we were able to use flat fields of a different observing run with the same grating and a slightly different central wavelength, $5570\text{\AA}$, to reduce the data. The other spectra, centered around $6083\text{\AA}$, are still reduced with the low S/N flat fields, introducing additional noise into the spectra.
However, the detector of the spectrograph does not show strong sensitivity variations along the spatial direction and variations along the dispersion direction are indirectly taken care of by the standard calibration procedure. Therefore, our analysis of the spectra will not be impacted.

\section{The ELQS-N Quasar Catalog}\label{sec_elqs_spring_catalog}

\begin{figure}
 \centering
 \includegraphics[width=0.5\textwidth]{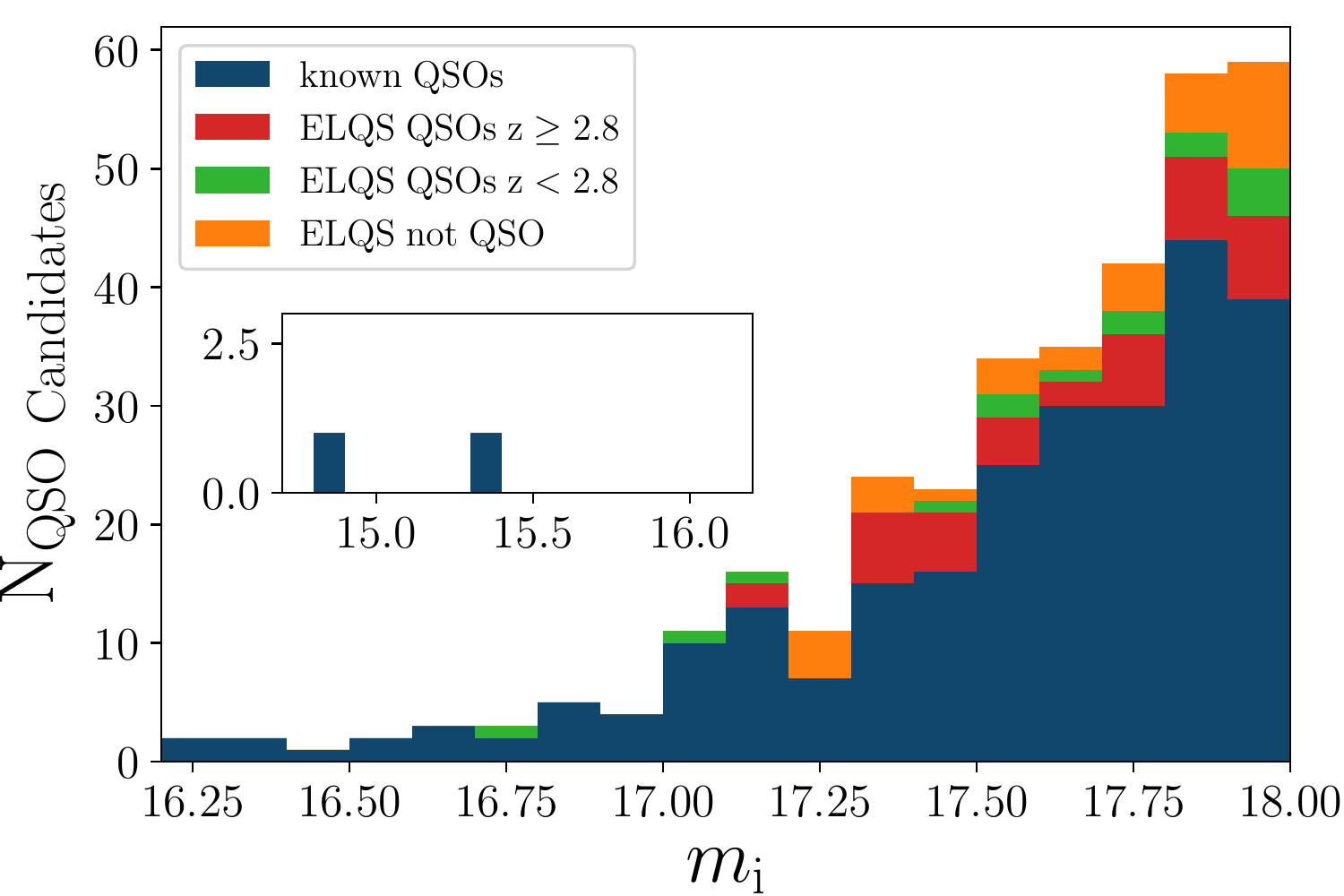}
 \caption{The distribution of the ELQS-N candidates as a function of apparent \textit{i}-band magnitude. Known quasars from the literature are shown in blue. Red and green refer to newly identified quasars with $z\geq2.8$ and $z<2.8$, respectively. Candidates that are not identified to be quasars are shown in orange. All candidates have been followed up.}
 \label{fig_elqs_spring_sample_dist}
\end{figure}

\begin{figure}[htb]
 \centering
 \includegraphics[width=0.5\textwidth]{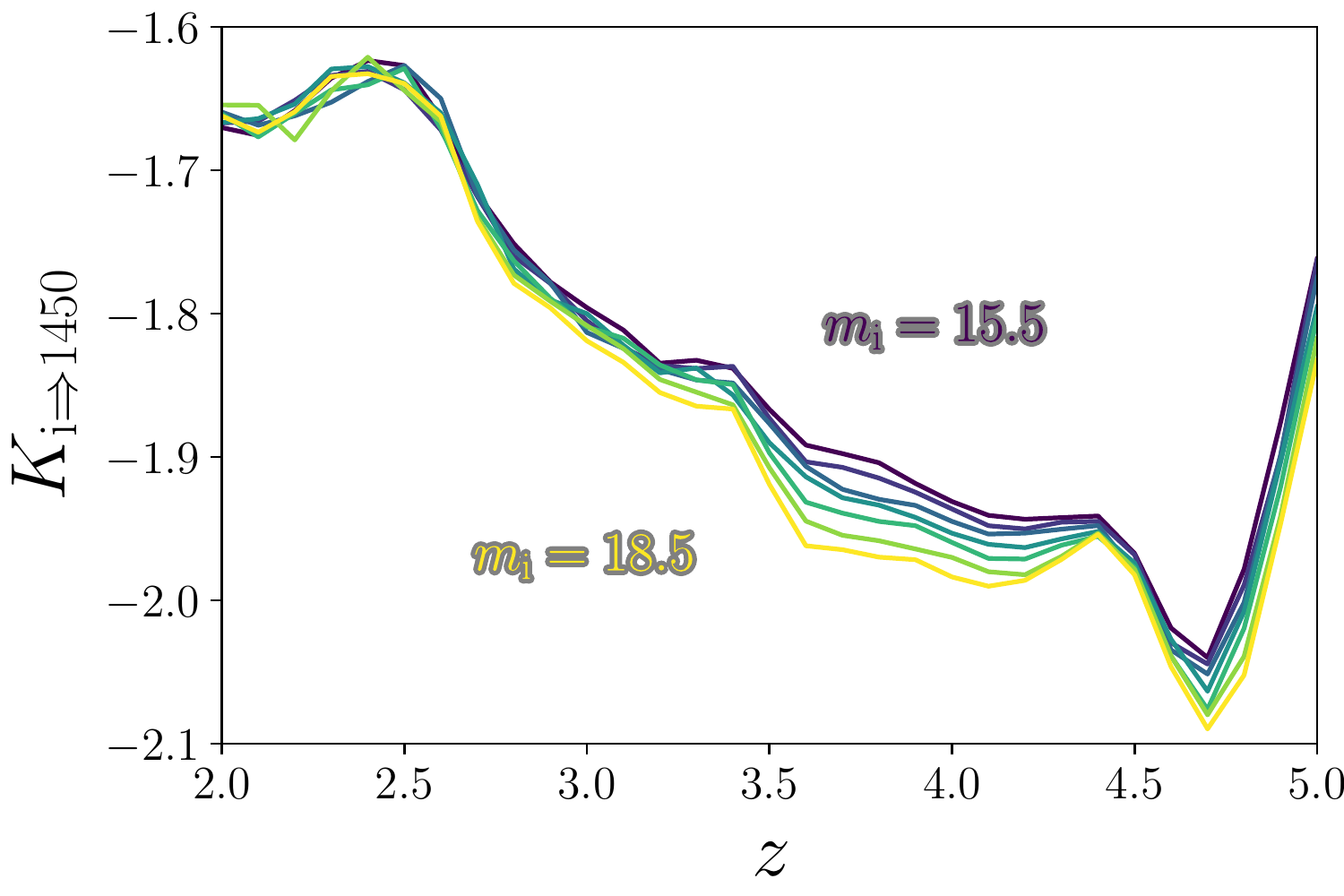}
 \caption{The luminosity-dependent k-correction ($K_{\rm{i}\Rightarrow 1450}$) as a function of redshift. It is derived from a large sample of simulated quasar spectra. The lines display the k-correction from the observed SDSS \textit{i}-band magnitude to $M_{1450}$ for $m_{i}=15.5$ (top) to $m_{i}=18.5$ (bottom), in steps of $\Delta m_{i}=0.5$.}
 \label{fig_kcorrection}
\end{figure}

\begin{figure}[htb]
 \centering
 \includegraphics[width=0.5\textwidth]{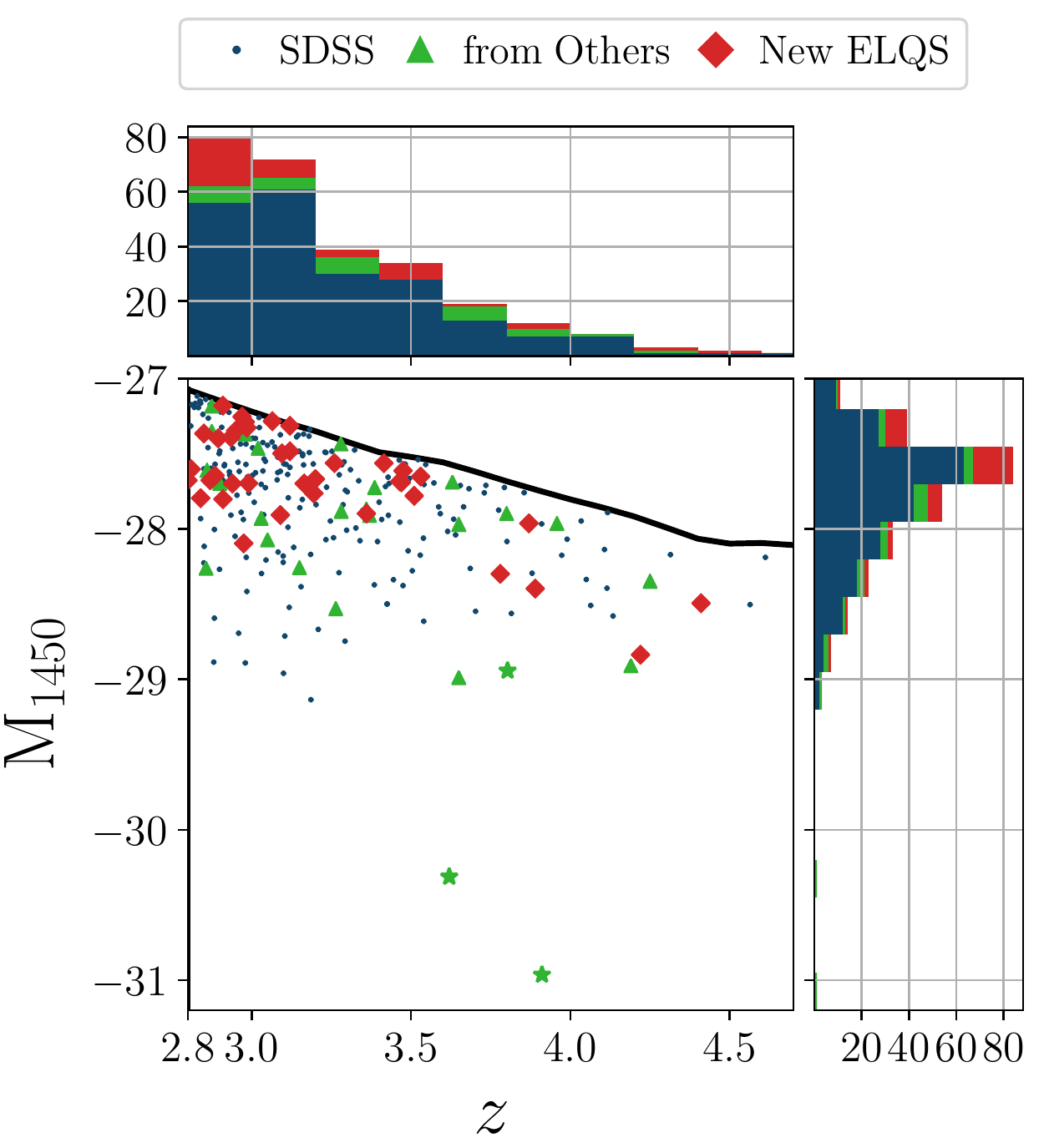}
 \caption{The distribution of all quasars in the ELQS-N sample in the $M_{1450}-z$ plane. Quasars from the SDSS DR7Q, DR14Q and with SDSS spectroscopy are shown as blue dots and labeled ``SDSS'', other quasars from the literature (MQC, Yang et al. quasar sample) are shown in green (triangles, stars) and the newly discovered quasars of this work are shown with red diamonds. We also show the distribution of the ELQS-N sample in histograms along the absolute magnitude ($M_{1450}$) and redshift ($z$) axis. The three green stars are the well-known quasar lenses  Q1208+1011, B1422+231B and APM 08279+5255.}
 \label{fig_elqs_spring_distribution}
\end{figure}
 
\begin{figure*}[htb]
 \centering
 \includegraphics[width=0.9\textwidth]{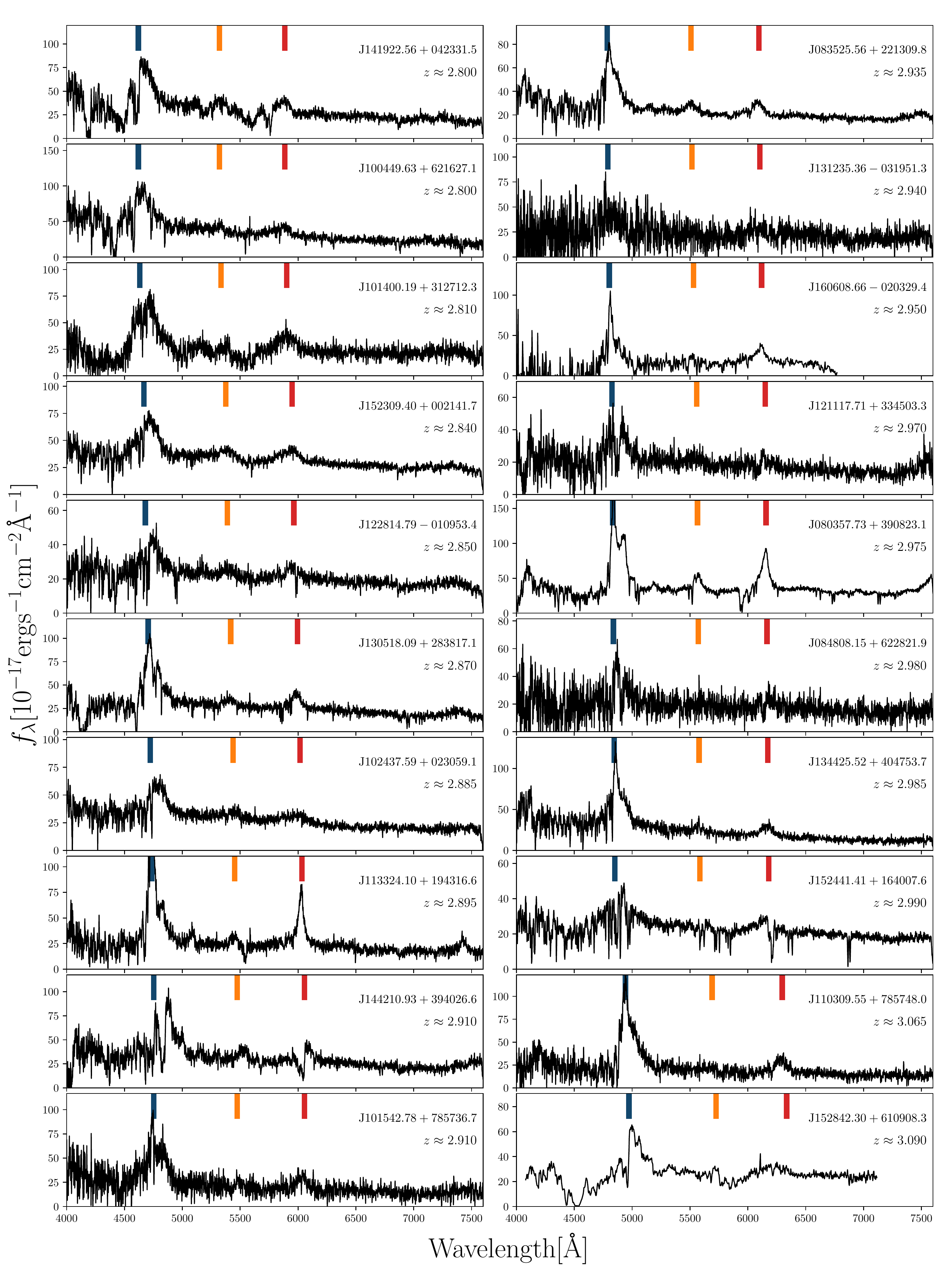}
 \caption{The discovery spectra of the newly identified ELQS-N quasars sorted by spectroscopic redshift. The  dark blue, orange and red bars denote the center positions of the broad $\rm{Ly}\alpha$, \ion{Si}{4} and \ion{C}{4} emission lines according to the spectroscopic redshift.}
 \label{fig_newqso_spectra}
\end{figure*}
 
\begin{figure*}[htb]
 \centering
 \includegraphics[width=0.9\textwidth]{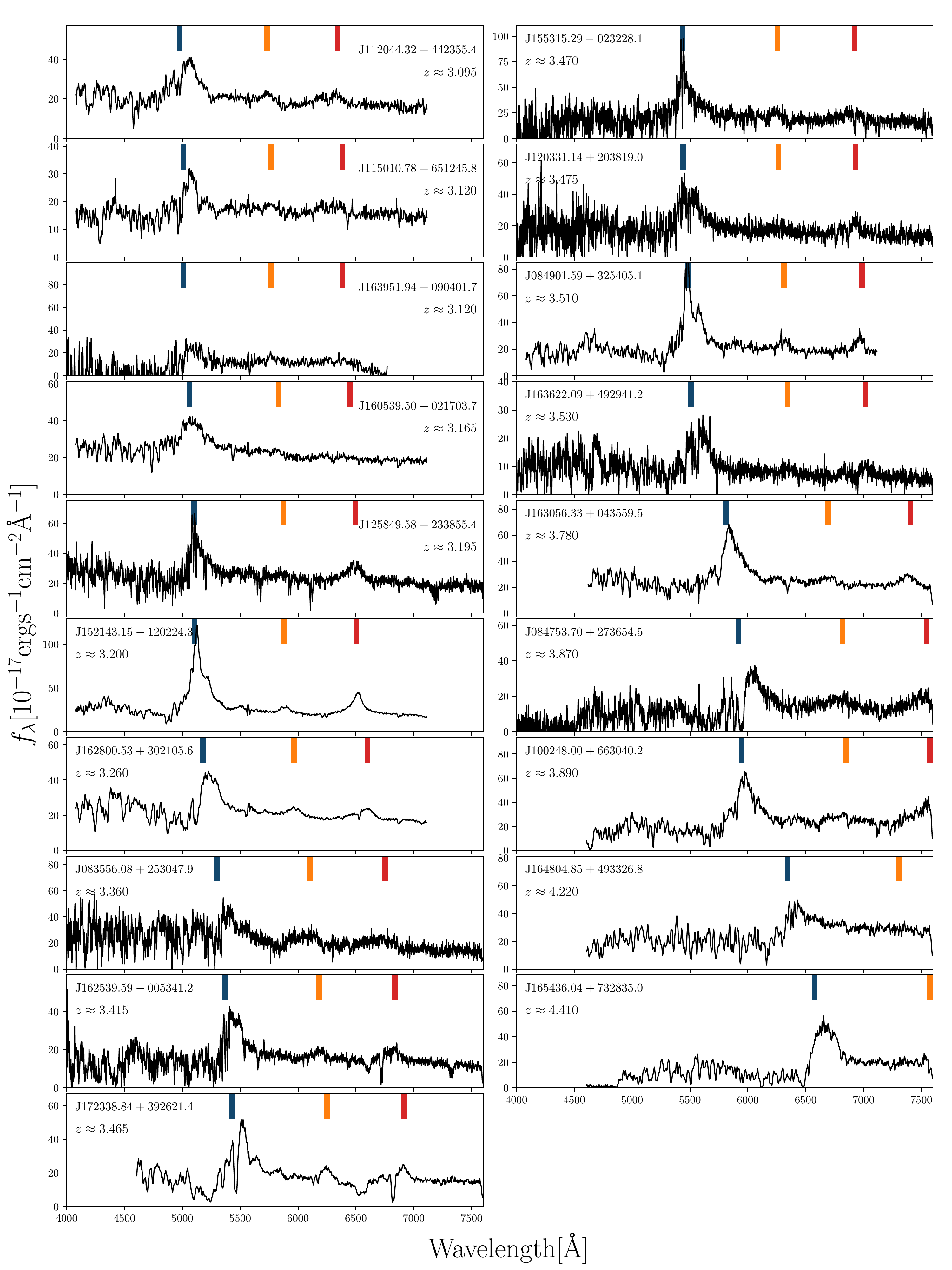}
 \caption[]{Continued from Figure\,\ref{fig_newqso_spectra}.}
\end{figure*}

\begin{table*}
\centering
 \caption{Newly discovered quasars at $z\geq2.8$ in the ELQS-N sample}
 \label{tab_elqs_spring_newqsos}
 \begin{tabular}{ccccccccc}
  \tableline
  \tableline
 R.A.(J2000) & Decl.(J2000) & $m_{i}$ &  $M_{1450}$ & Spectroscopic & Near-UV\tablenotemark{a} & Far-UV\tablenotemark{a} & BAL flag\tablenotemark{b} & Notes\tablenotemark{c} \\

 (hh:mm:ss.sss) & (dd:mm:ss.ss) & (mag)  & (mag) &  Redshift & (mag) & (mag) & & \\
  \tableline
08:03:57.742 & +39:08:23.05 & $17.10\pm0.01$ & -28.09 & 2.975 &- & - & 1 & 161218 \\
08:35:25.575 & +22:13:09.71 & $17.78\pm0.01$ & -27.39 & 2.935 &- & - & 0 & 170417 \\
08:35:56.094 & +25:30:48.05 & $17.57\pm0.02$ & -27.90 & 3.360 &- & - & 0 & 160412 \\
08:47:53.690 & +27:36:54.59 & $17.77\pm0.02$ & -27.96 & 3.870 &- & - & 0 & 170418\tablenotemark{d} \\
08:48:08.164 & +62:28:21.93 & $17.91\pm0.02$ & -27.29 & 2.980 &- & - & 0 & 170404 \\
08:49:01.620 & +32:54:05.51 & $17.75\pm0.01$ & -27.78 & 3.510 &- & - & 0 & 170517 \\
10:02:48.008 & +66:30:40.27 & $17.36\pm0.02$ & -28.40 & 3.890 &- & - & 0 & 170517\tablenotemark{d} \\
10:04:49.630 & +62:16:27.21 & $17.39\pm0.01$ & -27.68 & 2.800 &- & - & 0 & 161119 \\
10:14:00.201 & +31:27:12.33 & $17.46\pm0.01$ & -27.60 & 2.810 &- & - & 0 & 160311 \\
10:15:42.805 & +78:57:36.79 & $17.98\pm0.02$ & -27.18 & 2.910 &- & - & 0 & 170404 \\
10:24:37.592 & +02:30:59.24 & $17.48\pm0.02$ & -27.64 & 2.885 & $20.49\pm0.05$ & - & 0 & 160310 \\
11:03:09.562 & +78:57:47.98 & $17.98\pm0.01$ & -27.28 & 3.065 &- & - & 0 & 170404 \\
11:20:44.331 & +44:23:55.41 & $17.79\pm0.01$ & -27.50 & 3.095 &- & - & 0 & 170517 \\
11:33:24.095 & +19:43:16.71 & $17.74\pm0.01$ & -27.40 & 2.895 & $21.16\pm0.23$ & - & 0 & 170417 \\
11:50:10.783 & +65:12:45.79 & $17.82\pm0.02$ & -27.48 & 3.120 &- & - & 0 & 170517\tablenotemark{d} \\
12:03:31.145 & +20:38:18.69 & $17.90\pm0.02$ & -27.61 & 3.475 & $21.34\pm0.40$ &  $22.07\pm0.52$ & 0 & 170404 \\
12:11:17.716 & +33:45:02.65 & $17.94\pm0.02$ & -27.25 & 2.970 &- & - & 0 & 170404 \\
12:28:14.791 & -01:09:53.50 & $17.75\pm0.01$ & -27.36 & 2.850 & $22.13\pm0.26$ & - & 0 & 170405\tablenotemark{d} \\
12:58:49.601 & +23:38:55.54 & $17.58\pm0.02$ & -27.76 & 3.195 &- & - & 0 & 170504 \\
13:05:18.099 & +28:38:17.12 & $17.44\pm0.02$ & -27.67 & 2.870 &- & - & 0 & 160311\tablenotemark{d} \\
13:12:35.368 & -03:19:51.44 & $17.48\pm0.02$ & -27.70 & 2.940 &- & - & 0 & 170404\tablenotemark{d} \\
13:44:25.527 & +40:47:53.79 & $17.88\pm0.02$ & -27.32 & 2.985 &- & - & 0 & 170406 \\
14:19:22.545 & +04:23:31.75 & $17.46\pm0.02$ & -27.60 & 2.800 &- & - & 1 & 170404\tablenotemark{d} \\
14:42:10.930 & +39:40:26.69 & $17.36\pm0.02$ & -27.80 & 2.910 &- & - & 1 & 170404 \\
15:21:43.156 & -12:02:24.53 & $17.68\pm0.01$ & -27.67 & 3.200 & $21.57\pm0.40$ & - & 0 & 170518 \\
15:23:09.415 & +00:21:41.73 & $17.31\pm0.01$ & -27.79 & 2.840 &- & - & 0 & 170405 \\
15:24:41.404 & +16:40:07.57 & $17.51\pm0.02$ & -27.69 & 2.990 &- & - & 1 & 170405 \\
15:28:42.296 & +61:09:08.29 & $17.37\pm0.01$ & -27.91 & 3.090 &- & - & 0 & 170517 \\
15:53:15.300 & -02:32:28.05 & $17.83\pm0.02$ & -27.69 & 3.470 &- & - & 0 & 170406 \\
16:05:39.507 & +02:17:03.82 & $17.62\pm0.01$ & -27.70 & 3.165 & $22.76\pm0.46$ & - & 0 & 170518 \\
16:06:08.671 & -02:03:29.48 & $17.83\pm0.01$ & -27.35 & 2.950 &- & - & 0 & 150509 \\
16:25:39.599 & -00:53:41.10 & $17.94\pm0.01$ & -27.56 & 3.415 &- & - & 0 & 170405 \\
16:28:00.530 & +30:21:05.55 & $17.82\pm0.02$ & -27.56 & 3.260 &- & - & 0 & 170518 \\
16:30:56.335 & +04:35:59.42 & $17.39\pm0.01$ & -28.30 & 3.780 & $21.18\pm0.23$ &  $21.42\pm0.32$ & 0 & 170518\tablenotemark{e} \\
16:36:22.097 & +49:29:41.28 & $17.88\pm0.02$ & -27.65 & 3.530 & $22.02\pm0.26$ & - & 1 & 160313 \\
16:39:51.938 & +09:04:01.71 & $17.98\pm0.02$ & -27.31 & 3.120 &- & - & 0 & 150509\tablenotemark{d} \\
16:48:04.848 & +49:33:26.80 & $17.11\pm0.01$ & -28.84 & 4.220 &- & - & -1 & 170517 \\
16:54:36.038 & +73:28:35.04 & $17.58\pm0.02$ & -28.49 & 4.410 &- & - & -1 & 170517 \\
17:23:38.809 & +39:26:21.42 & $17.83\pm0.06$ & -27.68 & 3.465 &- & - & 1 & 170517 \\
\tableline
 \end{tabular}
\tablenotetext{1}{The near and far UV magnitudes were obtained from cross-matches within $2\farcs0$ to the \textit{GALEX} GR6/7 data release}
\tablenotetext{2}{Visual qualitative BAL identification flag: $1=$BAL; $0=$no BAL; $-1=$ insufficient wavelength coverage or inconclusive archival data}
\tablenotetext{3}{This column shows the observation date (YYMMDD) and provides further information on indidivdual objects.}
\tablenotetext{4}{These objects were also independently discovered by Yang et al.}
\tablenotetext{5}{See also \textit{HST} GO Proposal 13013 (PI: Gabor Worseck) and \citet{Zheng2015}}
\end{table*}

We have finished the spectroscopic identification campaign of ELQS-N, while data reduction of the ELQS-S still continues. With regard to the full ELQS candidate sample, at the time of this writing (2018 May), we have identified over 100 new quasars at $z\geq2.8$, of which 39 are located in the North Galactic Cap.

The ELQS-N quasar sample includes 375 candidates with $m_{i}\leq18.0$ and $z_{\rm{reg}}\geq2.8$ selected using our quasar selection algorithm (Section\,\ref{sec_selection_lit}). However, we discarded 35 candidates during a visual inspection of the photometry. All objects that were strongly blended in the \textit{WISE} bands or showed photometric artifacts, for instance, bright trails identified as the source, were not followed up spectroscopically.
Of the remaining 340 candidates 231 are known quasars from the literature at $z>2.8$, while 21 are low redshift ($z<2.8$) quasars (DR14Q: 13 objects, MQC: 5 objects, SDSS spectrum: 2 objects, DR7Q:1 object).
We have followed up all of the 88 candidates with optical spectroscopy and discovered 39 new quasars at $z\geq2.8$ and 15 quasars at $z<2.8$. The majority of our contaminants could be identified as K-dwarf stars, which have similar optical colors to quasars in our targeted redshift range.
Table\,\ref{tab_cand_sample} provides an overview of the candidate samples. The final ELQS-N quasar catalog includes 270 quasars with $m_{i}\leq18.0$ and $z\geq2.8$:

\begin{enumerate}
 \item 204 quasars from the DR14Q
 \item 39 newly identified quasars
 \item 15 quasars from the MQC
 \item 10 quasars from Yang et al. (2018, in preparation)
 \item 1 quasar with SDSS DR14 spectrum (BAL)
 \item 1 quasar from the DR7Q
\end{enumerate}

Excluding the 35 objects with unreliable photometry, we have identified 270 quasars out of a candidate sample of 340, resulting in a selection efficiency of $\sim79\%$.
In Figure\,\ref{fig_elqs_spring_sample_dist} we show a histogram of all 340 objects included in the original candidate catalog. All previously known quasars (regardless of redshift) are shown in blue. Quasars identified as part of the ELQS are colored red ($z\geq2.8$) and green ($z<2.8$). The remaining candidates, which were observed but identified not to be quasars, are shown in orange.

The majority of spectra cover the Ly$\alpha$, \ion{Si}{4} and \ion{C}{4} broad emission lines with a reasonable S/N. Only two of our quasar spectra at $z>4.0$ have just the Ly$\alpha$ line available. However, both spectra clearly show the Ly$\alpha$-forest blue-ward of the broad Ly$\alpha$ emission line. All ELQS-N quasars are type I quasars with broad emission lines (FWHM$> 1000\,\rm{km}{s}^{-1}$).
It is known that the \ion{C}{4} line can show offsets from the systemic redshift, which are correlated with quasar properties such as luminosity or radio loudness \citep[e.g.,][]{Richards2011}.
Hence, we determine the redshift of our discovered quasars based on visually matching a quasar template spectrum \citep{VandenBerk2001} to the observed spectra. We estimate that this method results in an uncertainty of $\Delta z\approx0.02$, which is accurate enough for the calculation of the QLF. 

We use a grid of simulated quasar spectra (Section\,\ref{sec_simqso}) to derive k-corrections as a function of redshift and magnitude to calculate the absolute monochromatic magnitude at rest frame $1450\,\text{\AA}$ ($M_{1450}$) from the SDSS \textit{i}-band magnitude. We interpolate on this grid to retrieve individual k-corrections for all quasars in our sample.
Figure\,\ref{fig_kcorrection} shows the luminosity-dependent k-correction ($K_{\rm{i}\Rightarrow 1450}$) calculated from the simulated spectra over a redshift range of $z=2-5$. 
We use it to account for the emission line contribution of \ion{C}{3}], \ion{C}{4}, \ion{Si}{4} and Ly$\alpha$ to the k-correction, which pass through the SDSS \textit{i}-band at redshifts $z=2.8-4.5$.

Figure\,\ref{fig_elqs_spring_distribution} shows the distribution of all quasars included in the ELQS-N catalog in the $M_{1450}$ and redshift plane. The newly discovered quasars are shown as red diamonds, whereas known quasars from the DR7Q, DR14Q and SDSS DR14 are depicted as small blue-filled circles and quasars from the other sources (MQC and Yang et al. 2018, in preparation) are shown as green triangles. 
The three green stars are the well-known quasar lenses Q1208+1011 ($z=3.8$) \citep{Bahcall1992, Magain1992}, B1422+231B ($z=3.62$) \citep{Patnaik1992}, and APM 08279+5255 ($z=3.91$) \citep{Ibata1999} that were missed by SDSS but included in our selection, as we discussed in \citet{Schindler2017}. A match of our ELQS-N catalog to all known quasar lenses registered in the NED, the CfA-Arizona Space Telescope LEns Survey of gravitational lenses (CASTLES; C.S. Kochanek, E.E. Falco, C. Impey, J. Lehar, B. McLeod, H.-W. Rix)\footnote{\url{https://www.cfa.harvard.edu/castles/}} and the SDSS Quasar Lens Search \citep{Inada2012} did not return additional results.

We display the spectra of all 39 newly identified quasars in the ELQS-N catalog in Figure\,\ref{fig_newqso_spectra}. It shows the fully reduced discovery spectra sorted by redshift, highlighting the $\rm{Ly}\alpha$, \ion{Si}{4}, and \ion{C}{4} emission lines with dark blue, orange and red bars at the top of each panel. 
In Table\,\ref{tab_elqs_spring_newqsos} we provide additional information for all of these objects, including the position in equatorial coordinates, SDSS apparent \textit{i}-band magnitude, the absolute magnitude at $1450\text{\AA}$, near- and far-UV magnitudes from the Galaxy Evolution Explorer (\textit{GALEX}), a flag indicating visual broad absorption line quasars classification and the determined spectroscopic redshift. 
The full ELQS-N catalog is available in a machine-readable format online. The catalog format is described in Table\,\ref{tab_elqs_cat_cols}.

\subsection{Matches to FIRST, GALEX, ROSAT 2RXS and XMMSL2}
We matched the ELQS-N quasar sample to the VLA Faint Images of the Radio Sky at Twenty-Centimeters (FIRST) catalog \citep{Becker1995} in an aperture of $3\farcs0$ to obtain $1.4\,\rm{GHz}$ peak flux densities. The cross match returned 34 matches, of which one, J083525.575+221309.71, is a newly discovered quasar at $z=2.94$. 
The FIRST survey area has been chosen to coincide with the SDSS North Galactic Cap footprint and therefore we can roughly estimate a radio-loud fraction (RLF) for the ELQS-N sample by counting all quasars with radio detections in FIRST. This results in an RLF of $\approx12.6\%$.

\citet{Ivezic2002} analyzed the radio fraction of quasars observed by the FIRST Survey and the SDSS. The authors argue that their sample does not show a dependency of the radio fraction on quasar luminosity or redshift. However, they state that dependencies on both could conspire to produce no observed effect.
\citet{Jiang2007} revisit the RLF of SDSS quasars at $z=0-5$ and $-30 \leq M_{i} < -22$. Contrary to \citet{Ivezic2002} the authors find that the RLF changes as a function of redshift and absolute magnitude. They find their discovery to be well fit by
\begin{equation}
 \log\left(\frac{\rm{RLF}}{1-\rm{RLF}}\right) = b_0 + b_z (1+z) + b_M (M_{2500}+26) \ ,
\end{equation}
where $b_0 = -0.132$, $b_z=-2.052$ and $b_M = -0.183$. 
As a result the RLF at $z=0.5$ declines from $24.3\%$ to $5.6\%$ as luminosity decreases from $M_{2500}=-26$ to $-22$ and at a fixed $M_{2500}=-26$ it decreases from  $24.3\%$ at $z=0.5$ to $4.1\%$ at $z=3$.

The ELQS-N sample has a median redshift of $z_{\rm{med}}\approx3.15$ and a median absolute magnitude of $M_{1450,\rm{med}}\approx -27.7$, which is equivalent to $M_{2500}\approx-28$, assuming a spectral slope of $\alpha=-0.5$. Using the relation of the RLF described in \citet{Jiang2007}, we predict an average RLF of $\sim8.5\%$.
Our measured RLF of $\approx12.6\%$ is in rough agreement with the predicted result using the \citet{Jiang2007} relation, taking into account that we classify quasars as radio-loud if they have any detection in FIRST. This likely increases our numbers in comparison to \citet{Jiang2007} who only considered quasars with a radio-to-optical ratio of $R = f_{6\rm{cm}}/f_{2500}>10$, where $f_{6\,\rm{cm}}$ and $f_{2500}$ are fluxes at the respective rest-frame wavelengths of $6\,\rm{cm}$ and $2500\,\text{\AA}$. We will revisit this analysis more rigorously in Paper III.

We further cross matched our catalog with the \textit{GALEX} GR6/7 Data Release \citep{Martin2005} in an aperture of $2\farcs0$, corresponding to the \textit{GALEX} position accuracy. We obtained photometry in the far- and near-UV bands at $1350-1750\,\text{\AA}$ and $1750-2750\,\text{\AA}$, where available. A total of 38 quasars have counterparts in GALEX, out of which 10 are detected in the far-UV band, 37 in the near-UV band, and 9 in both. Of the 39 newly discovered quasars, 8 have valid near-UV magnitudes and 2, J120331.145+203818.69 and J163056.335+043559.42, were detected in the far-UV as well.
The detection of luminous quasars in the near- and far-UV suggests that their flux was not fully absorbed by neutral hydrogen along the sight line of the quasar. These objects are prime targets to study the helium reionization of the universe \citep[see,][]{Worseck2011, Worseck2016}.

The rate of UV detections in the ELQS-N sample ($38/270\approx14\%$) is lower compared to the new ELQS quasars ($8/39\approx20\%$). A likely explanation is found in \citet{Worseck2011}. The authors confirm that the SDSS quasar selection preferentially selects quasars with intervening \ion{H}{1} Lyman limit systems at $3\lesssim z \lesssim 3.5$. The optical photometry of quasars at these redshifts without intervening \ion{H}{1} Lyman limit systems is bluer and shifts them into the stellar locus in optical color space.  
We have demonstrated \citep{Schindler2017} that our quasar selection overcomes the limitations of purely optical color selection by including near- to mid-IR photometry. It would not be surprising if it resulted in a higher rate of UV detections. However, our current sample is too small to be statistically relevant. Therefore, we will re-evaluate this fraction with the full sample.

Lastly, we match our quasar catalog with prematched AllWISE counterparts to X-ray detections \citep{Salvato2018} from the \textit{ROSAT} \citep{Truemper1982} reprocessed 2RXS catalog \citep{Boller2016} and the XMM Newton Slew 2 Survey (XMMSL2). These catalogs contain 106,573 counterparts to $0.1-2.4\,\rm{keV}$ 2RXS sources as well as 17,665 counterparts to $0.2-12\,\rm{keV}$ XMMSL2 sources. The AllWISE positions of our sample are matched to the AllWISE positions of the matched counterparts in a $6\arcsec$ aperture. We find 8 positional matches to the \textit{ROSAT} 2RXS catalog, all of which are already known quasars in the literature. The \textit{ROSAT} 2RXS fluxes are included in the ELQS-N quasar catalog.

\subsection{BAL Quasar Fraction}
We roughly estimate the fraction of BAL quasars  in the ELQS-N sample. First, we cross match our catalog to the SDSS DR12 quasar catalog \citep{Paris2017}, which flags all quasars with visually identified BAL features (\texttt{BAL\_FLAG\_VI}$=1$). This is a purely qualitative measure as opposed to the balnicity index (BI) \citep{Weymann1991} or the absorption index \citep{Hall2002}. The DR12Q BAL flag provides information on 190 of our 270 quasars, of which 40 are flagged as BALs. 
For all the remaining objects, we visually inspect their spectroscopy, where available, or use previous classification from the literature to determine their nature. Of our 39 newly discovered quasars 6 show BAL features (J080357.742+390823.05, J141922.545+042331.75, J144210.930+394026.69, J152441.404+164007.57, J163622.097+492941.28, and J172338.809+392621.42) and 2 do not have sufficient wavelength coverage to allow for clear classification.
In total we can identify 57 out of 262 ELQS-N quasars to be BALs, resulting in an observed BAL fraction of $\sim22\%$. 

Estimates of the \textit{traditional} BAL fraction, BAL quasars identified with $\text{BI}>0$, range from $10\%-30\%$ \citep[see e.g.,][]{Hewett2003, Trump2006, Dai2008, Maddox2008, Allen2011}. This large range arises in part from the difficulties in quantifying the selection of BALs accurately. Therefore, we limit comparisons to the observed traditional BAL fraction only. 

Within the SDSS DR3 quasar catalog \citet{Trump2006} found an observed traditional BAL fraction of $\sim10\%$ over a redshift range of $z=1.7-4.38$. However, quasar selection using near-infrared/infrared photometry is shown to result in quasar samples with a larger fraction of BAL quasars. \citet{Dai2008} demonstrated this on a 2MASS and SDSS DR3 matched quasar sample. This result was confirmed by \citet{Maddox2008}, who found a traditional observed BAL fraction of $\sim17.5\%$ in their \textit{K}-band excess selected quasar sample using the UKIRT (UK Infrared Telescope) Infrared Deep Sky Survey.\\
\citet{Allen2011} investigated the luminosity and redshift dependence of the BAL fraction using a quasar sample from SDSS DR6 and the SDSS DR5 quasar catalog \citep{Schneider2007}. The authors estimated an observed mean BAL fraction in \ion{C}{4} of $8\pm0.1\%$ and found that the intrinsic BAL fraction increases toward higher redshift. 
 
Our observed BAL fraction of $\sim22\%$ is higher in comparison to the studies of \citet{Trump2006}, \citet{Maddox2008}, and \citet{Allen2011}. Yet, it is unclear whether this is a result of our infrared based quasar selection \citep{Dai2008, Maddox2008}, a suggested redshift evolution of the BAL fraction \citep{Allen2011}, a luminosity dependence on the BAL fraction, or any combination of these effects.
A more detailed analysis of the balnicity and absorption index of our BAL quasar spectra is needed to draw further conclusions. However, this is beyond the scope of this work.

\subsection{Notes on individual objects}\label{sec_individ_objects}

\subsubsection{J083525.575+221309.71}
This quasar at $z=2.94$ was found to have a match in FIRST with a peak flux density of $F_{1.4\,\rm{,pk}}=1.05\pm0.145\,\rm{mJy}\,\rm{beam}^{-1}$ at a match distance of $0\farcs4$.

\subsubsection{J120331.145+203818.69}
This quasar at $z=3.475$ was matched to a \textit{GALEX} source with a distance of $0\farcs7$ with contributions in the near- and far- UV band. It has a near-UV magnitude of $21.34\pm0.40$ and a far-UV magnitude of $22.07\pm0.52$. However, there is a faint galaxy, J120331.19+203823.65, close to the quasar at a separation of $\approx5\arcsec$, which could contaminate the near- and far- UV flux measured by \textit{GALEX}.

\subsubsection{J163056.335+043559.42}
This quasar at $z=3.78$ also has \textit{GALEX} near- and far-UV counterparts with magnitudes of $21.18\pm0.23$ and $21.42\pm0.32$, respectively. The match distance between the SDSS and \textit{GALEX} coordinates is $\approx0\farcs8$. There is a faint point source, J163056.48+043602.55, with $i\approx22.46$ close to the quasar at a separation of $3\farcs84$ that could technically be confused with the quasar in \textit{GALEX}. Other nearby sources have separations of at least $10\arcsec$. This object was independently discovered by Gabor Worseck (see \textit{HST} GO proposal 13013) and further studied by \citet{Zheng2015}. It has not been formally published in any quasar catalog and we therefore include the object in the newly identified quasars of the ELQS-N sample.

\subsubsection{J131608.97+254930.43}
\begin{figure}
 \centering
 \includegraphics[width=0.5\textwidth]{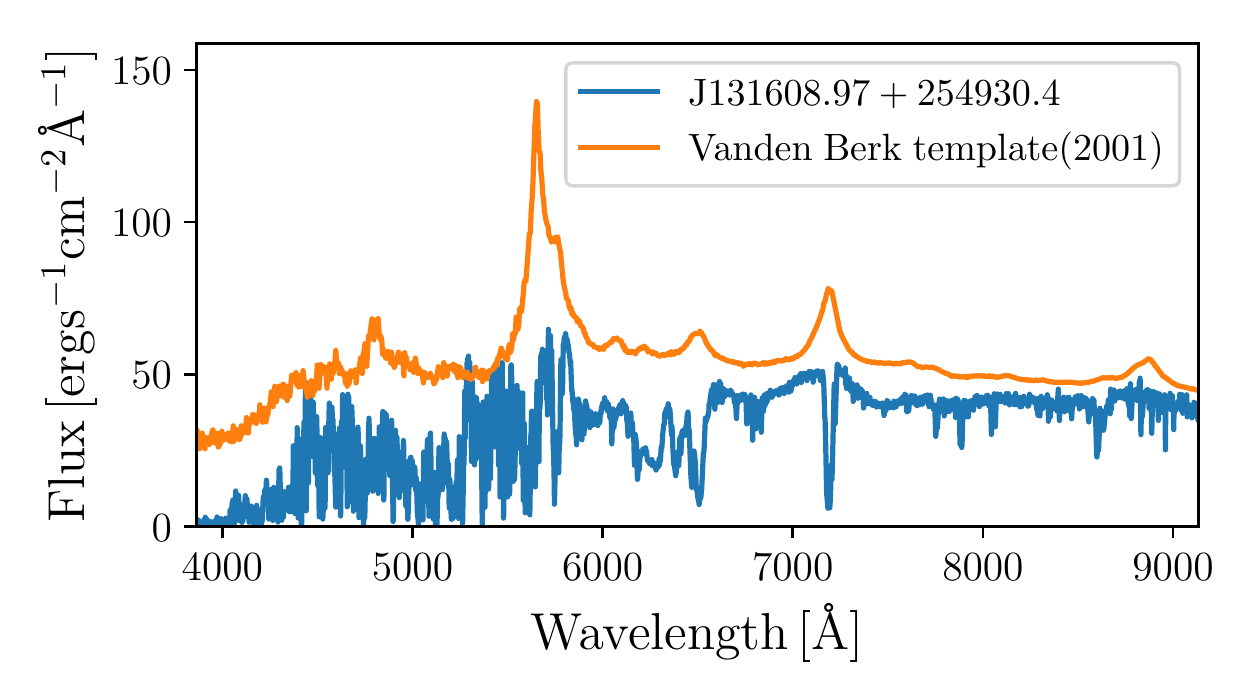}
 \caption{The SDSS spectrum of the low ionization BAL, J131608.97+254930.43, in comparison to the \citet{VandenBerk2001} quasar template shifted to a redshift of $z=3.65$.}
 \label{fig_sdss_spectrum}
\end{figure}

This object was observed as part of the Segue2 survey and was identified as an $z\approx3.24$ broad-line quasar by the pipeline. A visual re-evaluation of the pipeline spectrum resulted in the discovery that the object is a low ionization BAL quasar at $z\approx3.65$. We show its spectrum along with a quasar template for comparison in Figure\,\ref{fig_sdss_spectrum}.

\section{Selection function and completeness calculation}\label{sec_completeness}

To calculate the QLF for the ELQS sample, we have to carefully analyze the completeness of our quasar selection.
We utilize a large grid of simulated quasar spectra to model the quasar photometry and then apply our quasar selection to this simulated data set. We calculate the selection function for the full selection as well as for the photometric selection, the \textit{JKW2} color cut, quasar-star classification, and the redshift estimation individually.

\subsection{Simulated Quasar sample}\label{sec_simqso}
We constructed a sample of simulated quasar fluxes and magnitudes using \texttt{simqso}\footnote{\url{https://github.com/imcgreer/simqso}} (v1.1). \citet{McGreer2013} developed this software based on the spectral quasar model of \citet{Fan1999a}. It follows the assumption that quasar spectral energy distributions (SEDs) do not evolve with redshift \citep{Kuhn2001, Yip2004, Jiang2006}. 

The model of the quasar spectrum is based on a set of broken power laws ($f_\nu = A\cdot\nu^\alpha$). We model the quasar continuum using 11 break points and 12 mean slopes roughly based on the SED template of \citet{Elvis1994}. All break points and slopes are listed in Table\,\ref{tab_breakpoints_slopes}. 
The individual continuum slopes  for each spectrum are drawn from a Gaussian distribution around the mean slope ($\alpha$) with a width of $\sigma=0.3$.

Emission lines are then added onto the power-law continuum following the emission line profiles from the \citet{Glikman2006} template. The composite spectrum of fainter low redshift quasars seem to resemble the same emission line properties as high-redshift luminous quasars \citep{Selsing2016} and it is therefore suitable for our work.
The continuum and emission line model were chosen to be a good representation of observed flux ratios of SDSS quasars as a function of redshift (see Appendix\,\ref{app_flux_ratios}).

We have used the IGM absorption model of \citet{Worseck2011} to simulate $\rm{Ly}\alpha$ absorption along the quasar sight-lines.
The Fe emission is based on a template of \citet{Vestergaard2001}. 

In addition, \citet{Yang2016} included the Fe template of \citet{Tsuzuki2006} for the wavelength region of $2200-2500\rm{\AA}$ into the software, which separates \ion{Fe}{2} emission from the \ion{Mg}{2} $\lambda2897$ emission line. They further added Fe emission around $3500-7500\text{\AA}$ using the \citet{Boroson1992a} template.

For each band and survey the simulation includes photometric error models constructed from the observed magnitude dependent photometric error distributions. Using these models, photometric errors are added to the simulated photometry. These error models were previously extended by \citet{Yang2016} to encompass AllWISE \textit{W1} and \textit{W2} photometric bands. In order to conduct simulations for our survey, we have added the 2MASS \textit{J},\textit{H} and \textit{Ks} bands along with error models to the software.
The simulations do not include a model for intrinsic extinction of the quasar, since the main focus of this work is on the unobscured quasar population.

We calculated quasar photometry on a grid with 28 cells in a magnitude range of $14 < m_{i} < 18.5$ and 53 cells in a redshift range of $0.2 < z < 5.5$. Each cell was uniformly populated with 200 quasars, resulting in a total of 296,800 simulated quasars. $\rm{Ly}\alpha$ forest absorption was simulated with a total of 2000 different sight-lines up to $z=5.5$. 


\begin{table}[ht]
\normalsize
\centering
\caption{Wavelength break points and slopes for the simulated quasar spectrum}
\label{tab_breakpoints_slopes}
 \begin{tabular}{cc}
  \tableline
  Break Points in \AA & Slope before Break Point ($\alpha$) \\
  \tableline
  \tableline
  1250 & -1.50 \\
  2851 & -0.40 \\
  4809 & -0.20 \\
  8047 & -0.20 \\
  10,857 & 0.20 \\
  13,447 & -1.54 \\
  20,942 & -2.15 \\
  22,000 & -1.56 \\
  144,751 & -0.80 \\
  403,479 & -0.25 \\
  $>403,479$ & -1.05 \\
 \end{tabular}
\end{table}

\subsection{Completeness Analysis of the ELQS Candidate Selection}

In this section we evaluate the completeness of the ELQS in the SDSS footprint. This includes the photometric completeness as given by the AllWISE source catalog, the 2MASS PSC, and the SDSS DR13 data release (Panel (a) Figure\,\ref{fig_compl_comp}) and the completeness of our complex selection function. The latter is a combination of the \textit{JKW2} color cut (Panel (b) Figure\,\ref{fig_compl_comp}), the random forest classification (Panel (c) Figure\,\ref{fig_compl_comp}) and the random forest redshift estimation (Panel (d) Figure\,\ref{fig_compl_comp}). We calculate the completeness by applying our selection criteria to the grid of simulated quasars.

\subsubsection{Photometric Completeness} \label{sec_phot_compl}
Our photometric selection is based on the AllWISE source catalog rematched with 2MASS PSC, where we first apply quality criteria and the \textit{JKW2} color cut  resulting in a \textit{WISE}-2MASS all-sky catalog \citepalias[for details see][]{Schindler2017}. This catalog is subsequently matched to SDSS photometry within a $3\farcs96$ aperture. The influence of the \textit{JKW2} color cut will be discussed below, while we focus here on the photometric completeness only.

The AllWISE source catalog is complete to $95\%$ for point sources with magnitudes of $W1=17.1$ and $W2=15.7$, not taking into account losses due to source confusion or contamination. Confusion in the SDSS is known to decrease the overall source completeness for quasars down to $95\%$ \citep{Richards2006}. No quantitative analysis has constrained the effect of confusion to the overall completeness of point sources in the AllWISE source catalog. However, confusion in the AllWISE catalog will reduce the completeness to the SDSS value at minimum, due to the larger \textit{WISE} point spread function. Therefore, we adopt a uniform $95\%$ completeness level and the \textit{WISE} $95\%$ limiting magnitudes for the completeness calculation.
The AllWISE photometry is also known to have a variable characteristic sensitivity due to the variable depth of coverage. However, this only affects fainter magnitudes than are included in our selection due to our focus on bright sources ($m_{i}\leq18.0$).

The 2MASS PSC is uniformly complete to $J=15.8$, $H=15.0$, $K_{\rm{s}}=14.3$ (VEGA) at a level of $10\sigma$ photometric sensitivity. All sources have at least a $\rm{S/N}>7$ in one band or are detected in all three bands with a $\rm{S/N}>5$. 
However, the photometric sensitivity is a strong function of galactic latitude, where sources in the galactic plane are generally less complete due to source confusion in highly crowded fields. Furthermore, many more sources with fainter magnitudes are included in the 2MASS PSC still fulfilling their S/N criteria. For sources at high galactic latitudes and a photometric accuracy of $\rm{S/N}>5 (\sigma\leq0.198)$, the 2MASS PSC is roughly photometrically complete down to $J=16.8$, $H=16.1$, $K_{\rm{s}}=15.3$\footnote{See Figure\,7 on \url{https://www.ipac.caltech.edu/2mass/releases/allsky/doc/sec2_2.html}} (VEGA). We adopt these magnitude and error limits for our calculation of the photometric completeness and impose them for the QLF analysis.

We further add signal-to-noise ratio limits to the completeness calculation. They reflect the 2MASS PSC criteria above as well as our additional criteria on AllWISE sources, i.e. $\rm{S/N}>5$ in \textit{W1} and \textit{W2}.


Figure\,\ref{fig_compl_comp} a) shows the overall photometric completeness of the ELQS survey in the SDSS footprint. At magnitudes brighter than $m_{i}\leq17.5$ it is complete to $95\%$ at all redshifts. At fainter magnitudes the limiting depth of the 2MASS PSC is the reason for the strong decrease in the completeness, which falls below $20\%$ at $m_{i}\gtrsim18$.

\subsubsection{Completeness of the Quasar Candidate Selection}

The quasar candidate selection for the ELQS survey has three major components. Quasar candidates are selected by (1) passing the \textit{JKW2} color cut, (2) either are classified as quasars by the random forest method or are included in the \citet{Richards2002} high redshift color boxes and (3) having an estimated redshift of $z_{\rm{reg}}\geq2.8$. 

The \textit{JKW2} color cut is applied at the very beginning of the selection process, when we construct the \textit{WISE}-2MASS all-sky catalog of potential quasars from the AllWISE source catalog. Only after optical photometry from SDSS is matched to these sources, do we use random forests to estimate quasar redshifts and classify the candidates. To calculate the completeness, we use the same training set and hyper-parameters that were used in the construction of the candidate catalog \citepalias[see][]{Schindler2017}.
We quantify the completeness of these three criteria individually, to discuss their influence on the overall completeness of our survey.

\begin{figure*}
 \centering
 \includegraphics[width=\textwidth]{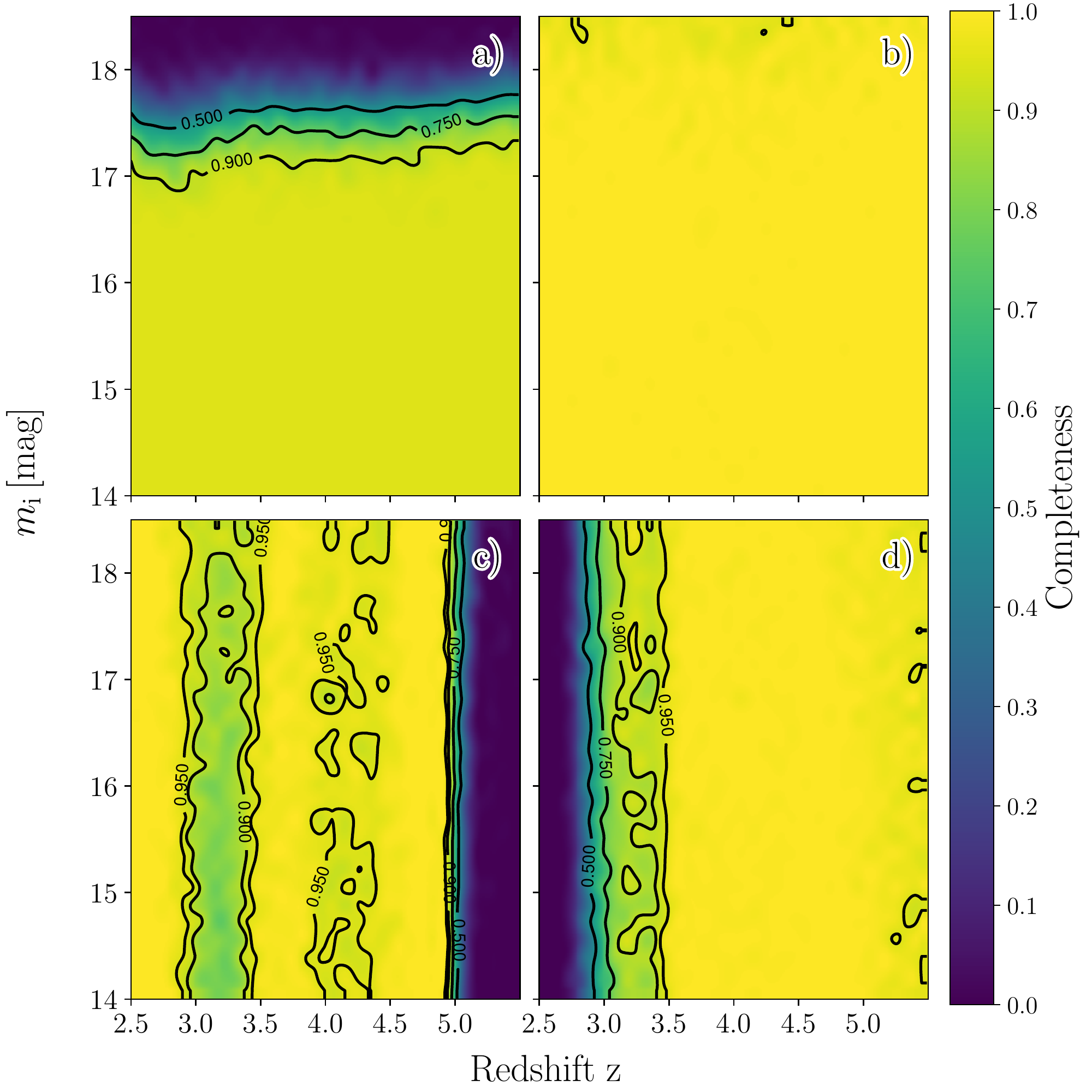}
 \caption{The ELQS completeness as a function of redshift and \textit{i}-band magnitude, as determined by the fraction of simulated quasars selected by our different criteria to all simulated quasars per grid cell: \textit{Panel a)} Photometric completeness of the AllWISE, 2MASS and SDSS catalogs. \textit{Panel (b)} Completeness of the \textit{JKW2} color cut. \textit{Panel (c)} Completeness of the quasar selection based on the random forest classifier and the high-redshift color inclusion regions. \textit{Panel (d)} Completeness of the redshift selection based on the random forest regression. Contour levels are drawn with solid lines at $50\%$, $75\%$, $90\%$, and $95\%$ completeness.}
 \label{fig_compl_comp}
\end{figure*}

Figure\,\ref{fig_compl_comp} (b) shows the completeness of the \textit{JKW2} color cut. It uniformly selects $>95\%$ of all quasars with  $m_{i}\leq18.5$. Only around $m_{i}=18.5$  the completeness seems to drop slightly below $95\%$. This demonstrates the very high completeness of the \textit{JKW2} color cut that we only evaluated in a $70\,\rm{deg}^2$ test region \citetalias{Schindler2017}.

The completeness of the selection based on the random forest classification and the inclusion regions is shown in Figure\,\ref{fig_compl_comp} (c). The completeness is generally independent of the i-band magnitude, but strongly dependent on the redshift. Around $z\approx5$ the completeness drops steeply from over $95\%$ to $0\%$. We attribute this behavior to poor training of the random forest method at these redshifts, because the empirical training set does not include any quasar with $m_{i}\leq18.5$ at $z>5$ and only 46 quasars with $z>5$ in total.
Between redshifts $z\approx3.0$ to $z\approx3.5$ quasar photometry is indistinguishable from stars in optical color space. The \textit{WISE} \textit{W1} and \textit{W2} bands help to lift this degeneracy and the completeness still reaches a level of $\sim80\%$. 
In a second redshift band from $z\approx4.0$ to $z\approx4.5$ the completeness decreases slightly below $95\%$. At these redshifts quasars cross the stellar locus in SDSS ugr and gri color space again.

\begin{figure}
 \centering
 \includegraphics[width=0.5\textwidth]{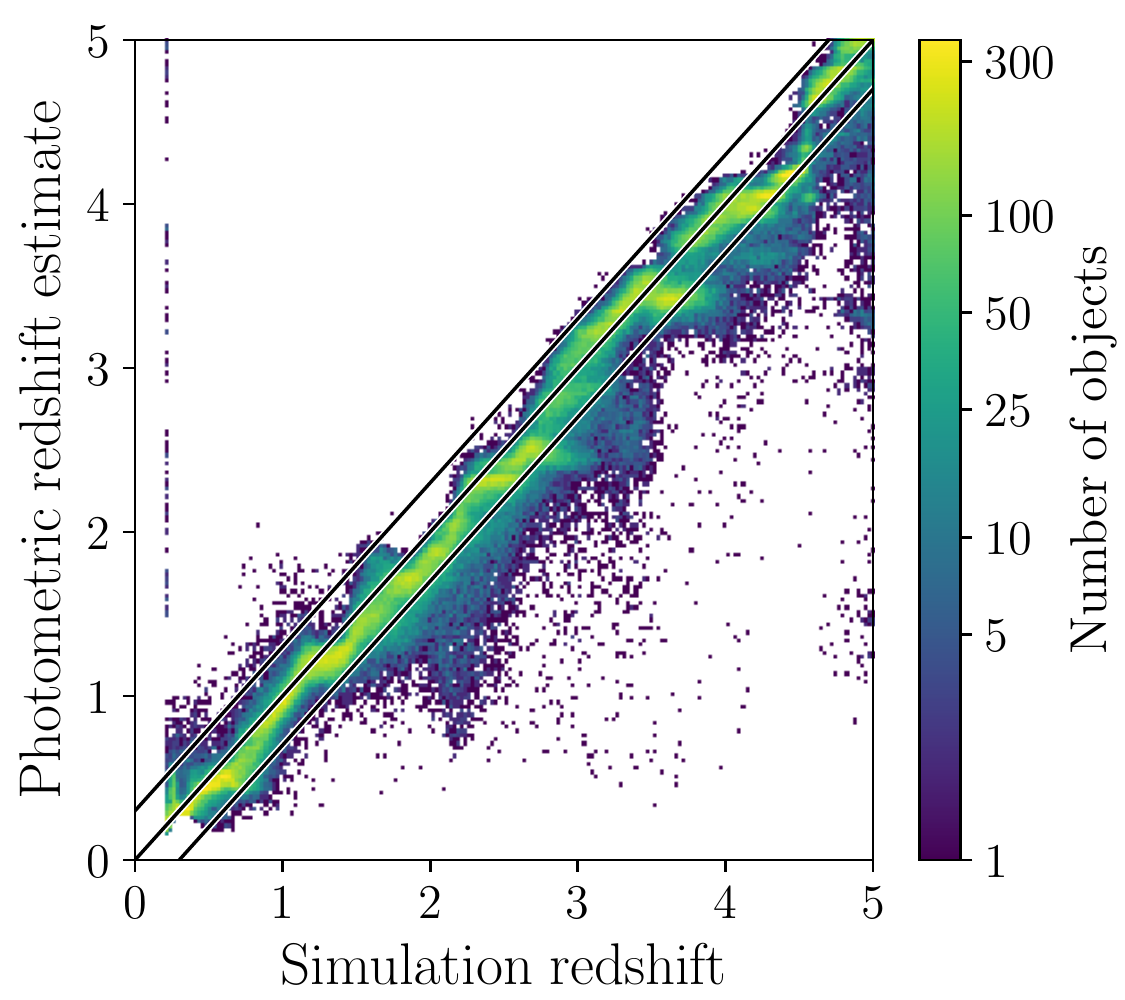}
 \caption{Photometric redshift estimates of the simulated quasar sample against their true redshifts. The majority of the sources lie within the two black lines around the diagonal ($\Delta z=0$), which illustrate the $|\Delta z|=0.3$ region. The data points are not distributed symmetrically around the diagonal. Especially at redshifts $z=1.8-3.5$ the majority of redshifts are underestimated.}
 \label{fig_sim_zz}
\end{figure}

Panel (d) in Figure\,\ref{fig_compl_comp} shows the completeness for our redshift selection based on the estimated redshift, $z_{\rm{reg}}$ that we calculated from random forest regression. Candidates with $z_{\rm{reg}}\geq2.8$ are selected and for redshifts higher than $z\approx3.5$, the completeness is well above $95\%$. Below the completeness drops steadily to $0\%$ at $z=2.8$.
If the estimated redshift $z_{\rm{reg}}$ was symmetrically distributed around the true redshift of the simulated quasars our completeness would extend to lower redshifts than $z=2.8$ and would also be higher. 
We have tested the random forest redshift regression in \citetalias{Schindler2017}. In that case, the test and the training set belonged to an empirical quasar sample built from the SDSS DR7Q and DR12Q quasar catalogs. The resulting analysis \citepalias[see][Figure\,7]{Schindler2017} showed that the regression redshifts are more symmetrically distributed around the spectroscopic redshifts of the test quasars.
This is clearly not the case if we train on the empirical quasar set and predict regression redshifts for the simulated quasar sample, as shown by the redshift estimates of the simulated quasars against their true values in Figure\,\ref{fig_sim_zz}.
This figure demonstrates that the simulated quasar sample is not a one-to-one representation of the empirical quasar training set. Our empirical training set might carry biases from the original SDSS and BOSS quasar selections that manifest as systematic deviations in the distribution of estimated to true redshifts in Figure\,\ref{fig_sim_zz}. For example, \citet{Worseck2011} showed that color-dependent biases in the original SDSS selection exist. They uncovered biases of the original SDSS quasar selection toward quasars with redder $u-g$ colors at $3.0\leq z\leq3.5$.
However, it is unclear at this point whether the empirical training set carries biases strong enough to induce these effects or if the photometric properties of the simulated quasar sample, dominated by the slopes and break points of the model spectrum, is misrepresenting reality.
Therefore, we will cautiously interpret our results at $2.8\leq z \leq3.0$, where completeness is most affected by the redshift estimation.

\subsubsection{Completeness of the ELQS in the SDSS footprint}

\begin{figure}[h]
 \centering
 \includegraphics[width=0.5\textwidth]{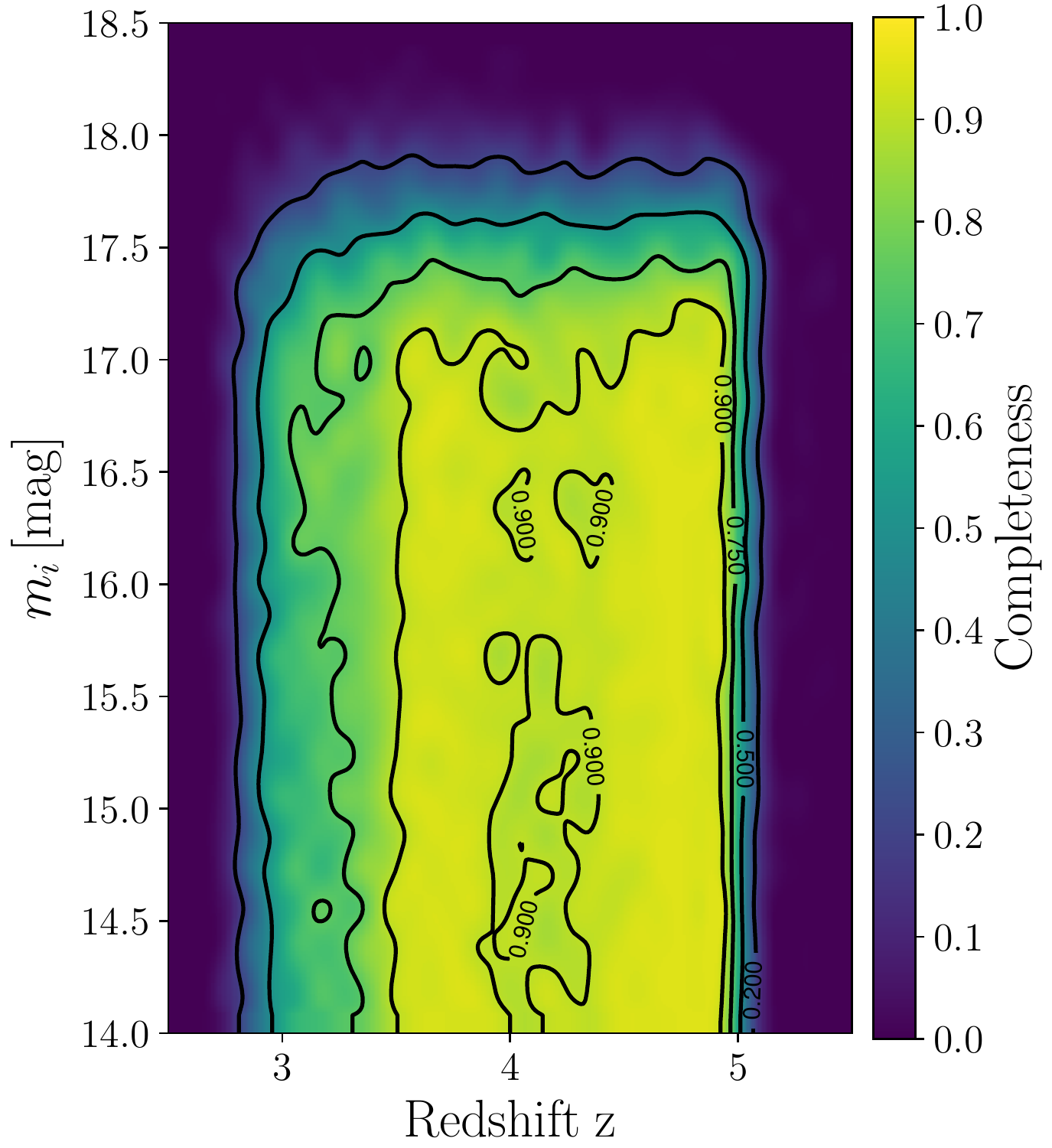}
 \caption{The completeness of the ELQS in the SDSS footprint as a function of redshift and \textit{i}-band magnitude, as determined by the fraction of simulated quasars selected by the full ELQS quasar selection to all simulated quasars per grid cell. Contour levels are drawn with solid lines at $20\%$, $50\%$, $75\%$, and $90\%$ completeness.}
 \label{fig_elqs_completeness}
\end{figure}

In Figure\,\ref{fig_elqs_completeness} we show the completeness of the full ELQS quasar selection as a function of redshift and \textit{i}-band magnitude. The shape and structure reflects the combination of the different selection criteria discussed above. 

In a region between $z\sim3.0$ and $z\sim5$ and at \textit{i}-band magnitudes brighter than $m_{i}\approx17.5$ the ELQS is complete to more than 70\%. Towards lower redshifts the completeness drops, because of our selection on the photometric redshift estimate ($z_{\rm{reg}}\geq2.8$). This results in a strip with 50\%-70\% completeness between $z\sim3$ to $z=3.5$ and $m_{i}<17.5$. 
The completeness drops considerably at $z\leq3.0$ due to the selection on regression redshifts, at $z\geq5.0$ due to the quasar classification selection, and at magnitudes fainter than $m_{i}\approx17.5$ due to the shallow photometry of the 2MASS PSC. 

As the completeness drops below $20\%$, the completeness corrections become large enough that the associated systematic uncertainties of the selection dominate. These are associated with the construction of the photometric catalogs from SDSS, 2MASS and AllWISE, and the random forest training sets and the assumptions of the model spectra.

%

The completeness of the SDSS quasar survey, which extends over a much wider redshift range than ELQS, is discussed in \citet[][their Figure\,6]{Richards2006}.
Unfortunately, a comparison between both surveys' completeness estimates is likely misleading, because the underlying quasar model, from which the selection function is derived, has evolved strongly over the last decade. This includes our choice of power-law slopes from which the SED is built, the inclusion of the Baldwin effect, a recent model of the Lyman-$\alpha$ forest, and updated Fe emission around the \ion{Mg}{2} emission line.
Since we recover known quasars from the literature not selected in the original SDSS selection and we are able to discover 39 new quasars missed by SDSS in the well-surveyed North Galactic Cap, it appears that our selection is generally more inclusive than the selection of the original SDSS quasar survey.

The SDSS BOSS quasar selection function is depicted in Figure 6 of \citet{Ross2013} and focuses on a fainter population of quasars ($m_{i}\geq18.0$). This is mirrored in their selection function as their estimated completeness drops from $\sim60\%$ at $m_{i}\approx18.0$ to below $1\%$ at $m_{i}\approx17.7$. In this case, there is very little overlap between the selection function of their and our survey, because we focus on the bright quasar population. As a result, it is not surprising that we were able to discover new bright quasars at $z=2.8-4.5$, which were not selected by BOSS.

\section{The ELQS-N QLF}\label{sec_lumfun}

%

In this section we attempt a first analysis of the ELQS QLF using a binned approach, a non-parametric analysis, and a maximum likelihood fit to the ELQS spring sample. Completeness corrections based on the selection function presented in the previous section are important for a consistent result.

Unfortunately, part of the ELQS-N quasar sample does not fully adhere to the uniform selection criteria of the 2MASS PSC that we adopted for our completeness calculation (see Section\,\ref{sec_phot_compl}). There is a range of objects with less stringent photometric criteria included in the \textit{WISE}-matched 2MASS PSC. For the sake of a clean analysis we enforce the discussed photometric criteria of the 2MASS PSC ($J\leq16.8$, $H\leq16.1$, $K_{\rm{s}}\leq15.3$ and $\rm{S/N}>7$ in one band or $\rm{S/N}>5$ in all three bands) on the ELQS-N sample. This reduces it from 270 to 120 quasars at $z\geq2.8$ (95 at $z\geq3.0$), of which 10 are newly identified and another 16 were not recovered in the SDSS DR14Q catalog. This reduced sample is only used for the QLF analysis. 
The ELQS-N catalog (see Appendix\,\ref{app_qso_catalog}), which contains the full ELQS-N quasar sample of 270 objects (last two rows of Table\,\ref{tab_cand_sample}), includes a boolean flag to indicate which objects were used in the QLF analysis in this Section.

\subsection{Binned QLF}\label{sec_binned_qlf}

The binned QLF is calculated on the ELQS-N QLF sample divided into discrete bins in redshift and absolute magnitude ($M_{1450}$). 
We construct four redshift and five magnitude bins with bin edges $z=2.8, 3.0, 3.5, 4.0, 4.5$ and $M_{1450}=-29.1, -28.7, -28.3, -28, -27.7, -27.5$.

The binned QLF is then calculated over the ELQS-N footprint \citepalias[$7,601.2\pm7.2\,\rm{deg}^2$, see][]{Schindler2017} using the $1/V_{\rm{a}}$ method \citep{Schmidt1968, Avni1980} with the modification of \citet{Page2000}. The incompleteness due to the selection function (see Sec.\,\ref{sec_completeness}) is corrected for the binned QLF calculation.

\begin{figure*}[htp]
 \centering
 \includegraphics[width=\textwidth]{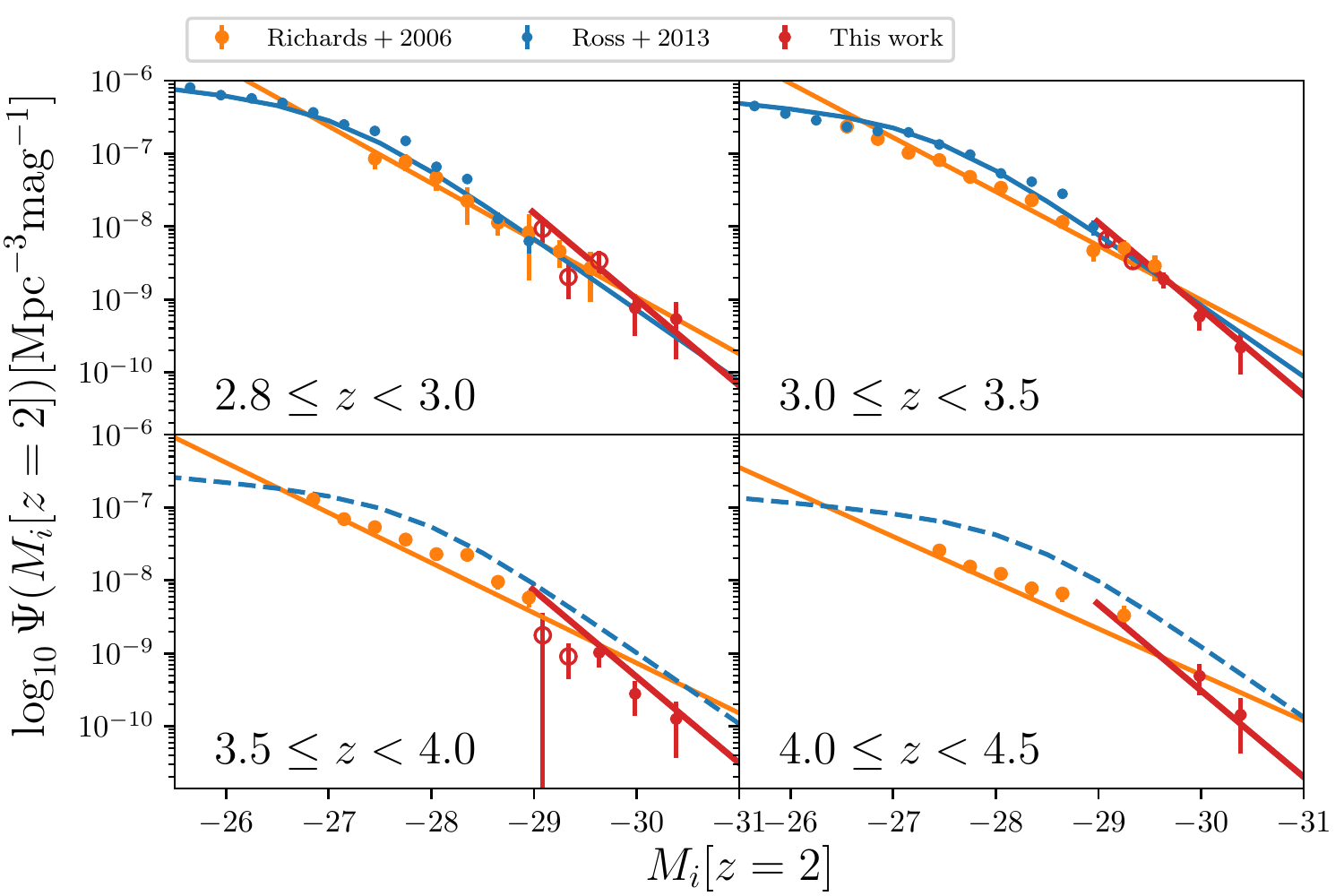}
\caption{The ELQS-N $m_{i}[z=2]$ QLF in the four redshift bins is shown in red. Open circles denote the data points that are either derived from unfilled bins or have an average completeness below $50\%$ ($N_{\rm{corr}}/N\geq2$). Previous estimates of the QLF are from the original SDSS DR3 \citep{Richards2006}, shown in orange, and from the BOSS DR9 \citep{Ross2013}, shown in blue.  The $1\sigma$ error bars show the purely statistical error due to the number of quasars per bin. The lines show parametric fits of the QLF to quasar distributions, where dashed lines indicate an extrapolation of the QLF prescription to higher redshifts. The red lines are from the maximum likelihood fit to the ELQS-N sample (Section\,\ref{sec_mlfit}, Table\,\ref{tab_mlqlffit} first row). The orange and blue lines correspond to the parametric fits to the QLF of \citep{Richards2006} and \citep{Ross2013}, respectively.}
 \label{fig_binnedqlf}
\end{figure*}

We display the resulting binned QLF in Figure\,\ref{fig_binnedqlf} in comparison to previous estimates of the optical QLF in SDSS \citep{Richards2006, Ross2013}. The absolute magnitudes $M_{1450}$ were converted to absolute \textit{i}-band magnitudes $m_{i}[z=2]\ (=M_{1450}-1.486)$ continuum k-corrected to a redshift of $z=2$ \citep{Richards2006}, to allow for better comparison to the previous publications.

The binned SDSS DR3 QLF \citep{Richards2006} and the SDSS DR9 QLF \citep{Ross2013} are shown with orange- and blue-filled circles, respectively. 
While we have chosen our three higher redshift bins to have identical boundaries compared to their work, our lowest redshift bin ($2.8 \leq z < 3.0$) differs slightly due to our redshift selection. Both the SDSS DR3 QLF and DR9 QLF do include quasars at $2.6 \leq z < 3.0$ in this bin.

The binned ELQS QLF is shown with filled and open red circles. The open circles denote values that are either in bins, which are not fully covered by our quasar selection, or have an average completeness below $50\%$ ($N_{\rm{corr}}/N\geq2$) and could therefore have substantial systematic biases due to the selection function. The error bars only show the purely statistical error based on the number of quasars per bin.

We further display the best fits to all three QLFs with solid lines colored corresponding to the binned data. The QLF fits are evaluated in the centers of the four redshift bins. The fit for the SDSS DR9 QLF is extrapolated beyond $z=3.5$, indicated by a dashed line. The best-fit values for the ELQS-N QLF are taken from the first row of Table\,\ref{tab_mlqlffit}.

With the ELQS survey, we are now able to extend the QLF by one magnitude toward the bright end. Our results of the binned QLF, shown in Figure\,\ref{fig_binnedqlf}, demonstrate that the bright-end slope is steeper than the bright-end slope of the \citet{Richards2006} QLF fit. Our QLF values agree with the QLF fit of \citet{Ross2013} in the two lower redshift bins, where their fit is not yet extrapolated. The values for the binned ELQS QLF are given Table\,\ref{tab_binnedqlf}.

\begin{table}
\normalsize
 \centering
 \caption{The binned QLF}
 \begin{tabular}{cccccc}
  \tableline
  $M_{1450}$ & $N$ & $N_{\rm{corr}}$ & $\log\Psi$ & $\sigma\Psi$ & bin filled \\
  \tableline
  \tableline
\multicolumn{6}{l}{$2.8\leq z <3.0$}\\
\tableline
-28.9 & 2 & 3.9 & -9.27 & 3.86 & True\\
-28.5 & 3 & 5.5 & -9.12 & 4.47 & True\\
-28.15 & 8 & 18.4 & -8.47 & 12.37 & True\\
-27.85 & 4 & 11.0 & -8.69 & 10.28 & True\\
-27.6 & 8 & 33.7 & -8.03 & 34.35 & True\\
\tableline
\multicolumn{6}{l}{$3.0\leq z <3.5$}\\
\tableline
-28.9 & 3 & 3.9 & -9.66 & 1.26 & True\\
-28.5 & 8 & 10.3 & -9.23 & 2.06 & True\\
-28.15 & 17 & 24.7 & -8.73 & 4.54 & True\\
-27.85 & 20 & 44.4 & -8.47 & 7.68 & True\\
-27.6 & 14 & 58.3 & -8.18 & 18.37 & False\\
\tableline
\multicolumn{6}{l}{$3.5\leq z <4.0$}\\
\tableline
-28.9 & 2 & 2.1 & -9.90 & 0.89 & True\\
-28.5 & 4 & 4.7 & -9.56 & 1.40 & True\\
-28.15 & 8 & 13.0 & -8.99 & 3.78 & True\\
-27.85 & 4 & 10.8 & -9.04 & 4.64 & False\\
-27.6 & 1 & 5.0 & -8.75 & 17.68 & False\\
\tableline
\multicolumn{6}{l}{$4.0\leq z <4.5$}\\
\tableline
-28.9 & 2 & 2.3 & -9.84 & 1.01 & True\\
-28.5 & 5 & 7.8 & -9.31 & 2.21 & True\\
\tableline
 \end{tabular}
 \label{tab_binnedqlf}
\end{table}

\subsection{Test of Correlation between the Luminosity and Redshift Distributions}\label{sec_corrtest}

The ELQS survey probes the extreme bright end of the quasar distribution at $z=2.8-4.5$. The resulting quasar sample is rather small and statistical uncertainties dominate the highest luminosity bins, as shown in Figure\,\ref{fig_binnedqlf}. 

However, if the redshift and luminosity distributions in the quasar sample are uncorrelated, we can calculate the marginal differential redshift and luminosity distributions. Because the distribution is marginalized over one variable, this approach allows for larger statistical samples to probe the quasar distribution along the other one.
The assumption that the data can be expressed in terms of two uncorrelated variables is identical to an underlying QLF of the form: $\Psi(M_{1450},z) = \rho(z)\psi(M_{1450})$. 


While at lower redshifts and luminosities the QLF is best fit by a double power law \citep[e.g.,][]{Boyle2000, Croom2004, Richards2006}, our survey only probes the highest luminosities.
Following \citet[][see their Figure\,19]{McGreer2013}, the break luminosity evolves from $M_{1450}^*\approx -25.6$ at z=2.8 to $M_{1450}^*\approx -27.0$ at z=5.0. Therefore, the break of the double power-law distribution \textit{is not sampled} in our survey and we can safely assume that our data can be represented by a single power law.

In this subsection, we test whether the slope of this power law is independent of redshift using the correlation test for truncated samples developed by \citet{Efron1992} and \citet{Maloney1999}, which was generalized to arbitrary selection functions by \citet{Fan2001b}.
We quantified the selection function of the ELQS survey and display it in Figure\,\ref{fig_elqs_completeness}. Our selection function is not truncated at the bright end, but it does have a complicated structure and decreases continuously toward higher and lower redshifts as well as higher apparent magnitudes. Therefore, we follow the methodology laid out in \citet{Fan2001b} to perform the correlation test.
For this and the following section (Section\,\ref{sec_cminus}) we will omit the wavelength subscript on the absolute monochromatic magnitude $M_{1450}$ to make the mathematical expressions more clear. 

The probability of selecting a quasar with a certain magnitude $M$ and redshift $z$ is given by our selection function $p(M,z)$. 
At first we define the comparable or associated data set $J_i$ for each quasar $i$ in the sample,
\begin{equation}
J_i = \{ j: M_j < M_i \} \ . \label{eq_comp_set}
\end{equation}
It includes all quasars that have a lower absolute magnitude than the object $i$ in question. 
In the next step, we determine the total weighted number of objects in each set $J_i$ by weighting each point $j$ in the set. The weight is proportional to the selection probability $p(M_i,z_j)$ if the object had the same magnitude as the object in question (object $i$), for which the comparable set was determined. It is further inverse proportional to its own selection probability $p(M_j,z_j)$. We define the quantity $T_i$ for each set $J_i$ as a measure of the ``total weighted number'' of objects in the set $J_i$ selected by our selection function $p(M,z)$:
\begin{equation}
T_i = \sum_{j=1}^{N_i} \frac{p(M_i,z_j)}{p(M_j,z_j)} \ .
\end{equation}
Here $N_i$ is the actual number of quasars in the comparable set $J_i$.
In the next step we calculate the rank $R_i$ of redshift $z_i$ in the comparable set $J_i$:
\begin{equation}
 R_i = \sum_{j=1}^{N_i} \frac{p(M_i,z_j)}{p(M_j,z_j)} \ , \text{where}\ z_j<z_i \ ,
\end{equation}
where $T_i$ was the total summed weight of all objects in set $J_i$, $R_i$ is the total summed weight of the subset $\{j\in J_i; z_j<z_i\}$.
If the distribution of quasars is a separable function of redshift $z$ and absolute magnitude $M$ and hence those parameters are independent from each other, then the rank $R_i$ should be distributed uniformly between $0$ and $T_i$. In the assumption of a uniform distribution the expectation value $E_i$ and variance $V_i$ for each set should follow:
\begin{align}
 E_i&= \frac{T_i}{2} & V_i&=\frac{T_i^2}{12}
\end{align}
At last one can define a single statistic $\tau$:
\begin{equation}
 \tau = \frac{\sum_i(R_i-E_i)}{\sqrt{\sum_i V_i}}
\end{equation}
This statistic $\tau$ is equivalent to Kendell's $\tau$ statistic.
In the case of $|\tau|\lesssim1$, the luminosities and redshifts can be regarded as uncorrelated parameters at the $\sim1\sigma$ level and can be treated independently.
Using our selection function, we calculate the $\tau$ statistic for the ELQS-N sample. 
For the cosmology specified in the introduction, we obtain $\tau=-0.29$ ($\tau=0.0$; $M_{1450}\leq-27.7$) for $2.8\leq z \leq 4.5$ and $\tau=-0.80$ ($\tau=-0.91$; $M_{1450}\leq-27.7$) if we restrict the redshift range to $3.0\leq z \leq 4.5$. Therefore, we can safely assume that our data, limited in absolute magnitude and redshift by our survey design, can be represented as a bivariate function of two uncorrelated variables, absolute magnitude $M$ and redshift $z$.

\subsection{The Differential Marginal Luminosity Functions}\label{sec_cminus}
We have demonstrated that one can regard the absolute magnitude and redshift distributions of the ELQS quasar sample as uncorrelated. Therefore, one can write the bivariate luminosity function as $\Psi(M,z) = \rho(z)\psi(M)$, where $\psi(M)$ and $\rho(z)$ are the marginal distributions of the luminosity function in the redshift and the absolute magnitude directions.
Using the Lynden-Bell $C^{-}$ estimator \citep{LyndenBell1971} we can now calculate the cumulative marginal luminosity function along the magnitude (luminosity) direction $\Phi(M) \equiv \int \psi(M)\rm{d}M$:
\begin{equation}
 \Phi(M_j) = \Phi(M_1) \prod_{k=2}^{j} (1 + N_k^{-1}) \ .
\end{equation}
$N_k$ denotes the total weighted number of objects in the comparable set. For an arbitrary selection function we use the comparable set defined in equation\,\ref{eq_comp_set} and calculate
\begin{equation}
 N_k = \sum_j^{} \frac{p(M_k,z_j)}{p(M_j,z_j)} \ ,
\end{equation}
instead of the absolute number of objects in the comparable set. For this calculation all objects need to be sorted along the absolute magnitude direction: $M_1 < ...< M_{i-1} < M_i < ... M_N$.

In analogy to $\Phi(M_j)$ the cumulative marginal distribution in redshift is
\begin{equation}
 \sigma(z_j) = \sigma(z_1) \prod_{k=2}^{j}(1 + W_k^{-1}) \ ,
\end{equation}
where $W_k$ is now the summed weight of the comparable sample, defined by the equation\,\ref{eq_comp_set} for all $z_j < z_k$:
\begin{equation}
  W_k = \sum_j^{} \frac{p(M_j,z_k)}{p(M_j,z_j)} \ .
\end{equation}
In addition, all objects are sorted in order of their redshifts: $z_1 < ... < z_{i-1} < z_i < ... < z_N$.

In our case the $M$ and $z$ distributions are uncorrelated. However, if they were correlated, one has to transform the correlated variables $(M,z)$ to a set of uncorrelated ones $x(M,z),y(M,z)$. Then the $C^{-}$ estimator can be used to calculate their marginal distributions \citep{Maloney1999}.

The $C^{-}$ estimator provides the shape of the two marginal cumulative distributions, but not their normalization. Both distributions can be normalized by requiring that the total number of observed objects is equal to the predicted number:
\begin{equation} \label{eq_norm}
\begin{split}
 N_{\rm{obs}} &= \int_0^{\infty} \psi(M)\sigma(z_{\rm{max}}(M))\rm{d}M \\ 
 &= \int_0^{\infty} \rho(z)\Phi(M_{\rm{max}}(z))\rm{d}z \ . 
\end{split}
\end{equation}

We described how the $C^{-}$ estimator offers an efficient non-parametric method to calculate the marginal cumulative distributions without the need to bin the data. In order to derive the marginal differential distributions binning along one variable is still needed. In comparison to the $1/V_{a}$ method, where we binned the data in both the $M$ and $z$ directions, the $C^{-}$ estimator retains a larger data sample when calculating either marginal distribution.

We use a modified version of the $C^{-}$ estimator offered by the \texttt{astroML}\footnote{https://github.com/astroML} library  \citep[see also][]{Ivezic2014_DataMining} to calculate the normalized differential distributions in absolute magnitude $\psi(M_{1450})$ and redshift $\rho(z)$ with errors estimated on 20 bootstrap samples of our data. 

The marginal differential magnitude distribution $\psi(M)/\Phi(M\leq-27.7)$, the number density of quasars as a function of magnitude, was calculated in magnitude bins with edges $-27.7, -28, -28.3, -28.7,$ and $-29.1$ and normalized to the total number of observed object $N_{\rm{obs}}$ with $M\leq-27.7$. The bin edges were chosen to be identical to the binned QLF (Section\,\ref{sec_binned_qlf}). We have fit this distribution with a power law, $\log(\psi(M_{1450})) \propto - 0.4 \cdot (\beta+1) \cdot M_{1450}$, corresponding to $\psi(L)\propto L^{-\beta}$, to estimate the slope of the bright-end QLF. The maximum likelihood fit for all quasars with $M_{1450}\leq-27.7$ was calculated for two cases. Over the entire redshift range ($2.8\leq z \leq 4.5$) we find the best-fit slope to be  $\beta= -3.98\pm0.18$, while over a redshift restricted sample ($3.0\leq z \leq4.5$) we find $\beta=-4.21\pm0.23$.

\begin{figure*}[htp]
 \includegraphics[width=\textwidth]{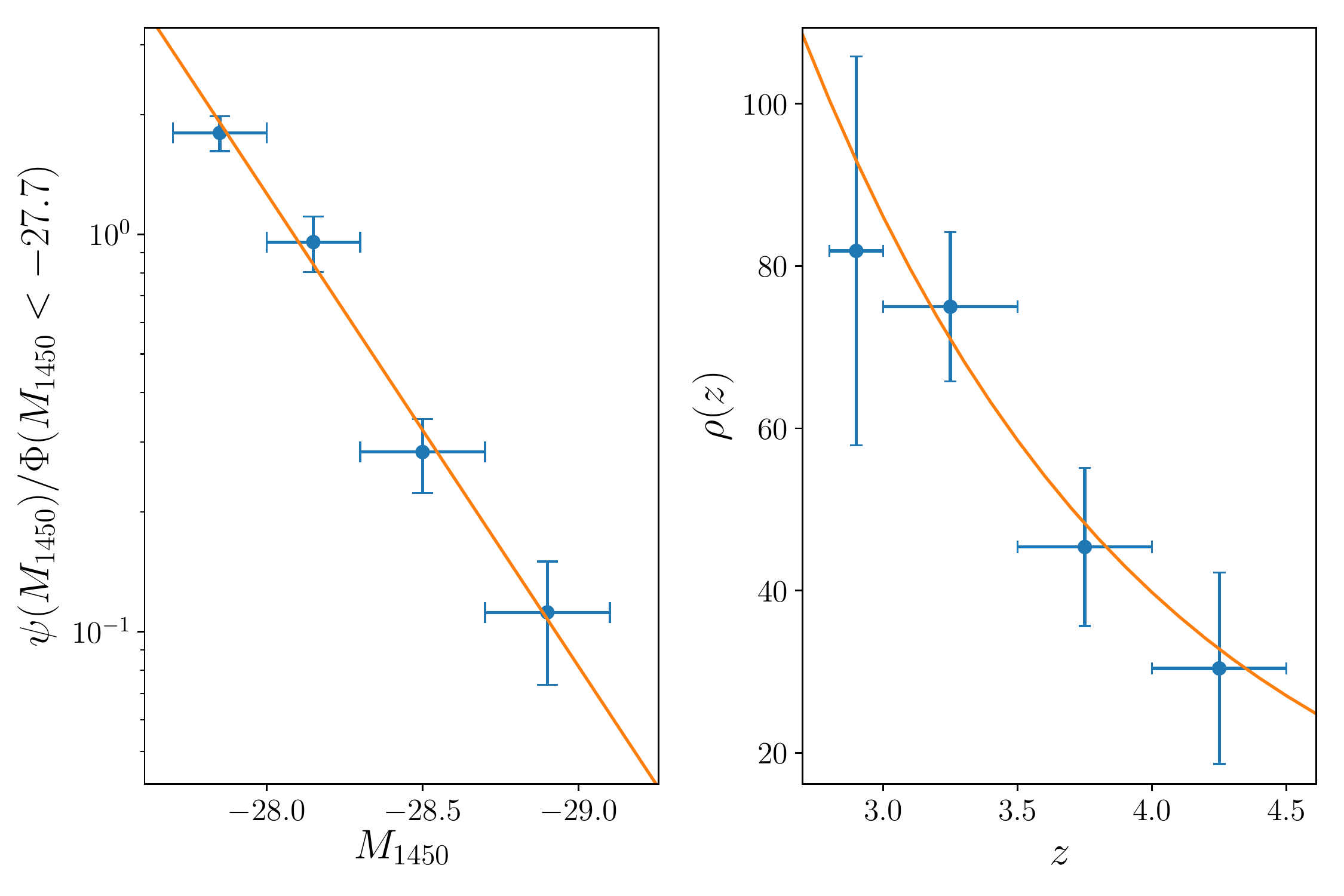}
 \caption{Left: the normalized marginal differential distribution of the QLF $\psi(M_{1450})/\Phi(M_{1450}\leq-27.7)$ as a function of absolute magnitude $M_{1450}$. The error bars in the magnitude direction represent the bin width, while the errors in $\psi(M_{1450})$ show the $1\sigma$ statistical error margins from the bootstrap sampling. The orange line is the maximum likelihood fit to the data points, $\psi(M_{1450})\propto 10^{-0.4(-3.98+1)\cdot M_{1450}}$. Right: the spatial density of the QLF $\rho(z)$, as a function of redshift $z$. The error bars in the redshift direction show the width of the redshift bins, while the error bars along $\rho(z)$ show the $1\sigma$ statistical error margins from the bootstrap sampling. The orange line is the maximum likelihood fit to the data points, $\log(\rho(z)) \propto -0.31\cdot z$.}
 \label{fig_marg_dist}
\end{figure*}

Similarly, we calculate the marginal differential redshift distribution $\rho(z)$, the spatial density of quasars as a function of redshift, over redshift bins with edges $2.8, 3.0, 3.5, 4.0,$ and $4.5$. We fit $\rho(z)$ with an exponential, $\log(\rho(z)) \propto \gamma \cdot z$, corresponding to $\rho(z)\propto 10^{\gamma z}$. The first fit includes all quasars with $2.8\leq z \leq 4.5$, while the second one only focuses on the objects with $3.0\leq z \leq4.5$. We find best-fit values of $\gamma=-0.31\pm0.02$ and $\gamma=-0.34\pm0.02$, respectively.
Both differential distributions as well as the best-fit power laws over the entire redshift range ($2.8\leq z \leq 4.5$) are shown in Figure\,\ref{fig_marg_dist}.

\subsection{Maximum Likelihood Estimation of the QLF}\label{sec_mlfit}

\begin{table*}[htb]
\normalsize
 \centering
 \caption{Maximum Likelihood Estimation Fit Parameters for the QLF}
 \label{tab_mlqlffit}
 \begin{tabular}{ccccc}
 z & $M_{1450}$ & $\log[\Psi^{\star}_0]$ & $\gamma$ &  $\beta$  \\
  \tableline
  \tableline
 2.8-4.5 & -31.5 to -27 & $-4.92^{+0.38}_{-0.37}$ & $-0.38^{+0.12}_{-0.12}$ & $-3.96^{+0.21(0.61)}_{-0.22(0.68)}$ \\
 \tableline
 3.0-4.5 & -31.5 to -27 & $-4.77^{+0.49}_{-0.47}$ & $-0.41^{+0.15}_{-0.15}$ & $-4.01^{+0.24(0.71)}_{-0.25(0.79)}$ \\
 \tableline
  2.8-4.5 & -31.5 to -28 & $-4.91^{+0.54}_{-0.53}$ & $-0.37^{+0.14}_{-0.14}$ & $-4.01^{+0.34(0.95)}_{-0.36(1.17)}$ \\
 \tableline
 3.0-4.5 & -31.5 to -28 & $-5.18^{+0.63}_{-0.64}$ & $-0.31^{+0.17}_{-0.17}$ & $-3.97^{+0.37(1.03)}_{-0.40(1.29)}$ \\
 \tableline
 \end{tabular}
\end{table*}

Instead of binning the data in redshift or magnitude, we derive parametric fits to the full data set using maximum likelihood estimation. The maximum likelihood estimate for the QLF $\Psi(M,z)$ can be calculated by minimizing the log likelihood function given by \citet{Marshall1983} as
\begin{equation}
\begin{split}
 S =& -2 \sum_i^N \ln\left(\Psi(M_i,z_i)(p(M_i,z_i)\right) \\
 & +2 \int \int \Psi(M,z) p(M,z) \frac{\rm{d}V}{\rm{d}z} \rm{d}M\rm{d}z \ . 
\end{split}
\end{equation}
The first term is the sum over all QLF contributions from the observed quasar sample multiplied by the selection probability $p(M,z)$ derived from the completeness calculation. The second term provides the normalization by integrating the QLF model $\Psi(M,z)$ over the entire redshift and magnitude range of the sample.
We derive the confidence interval from the likelihood function using a $\chi^2$ distribution in $\Delta S = S-S_{\rm{min}}$ \citep{Lampton1976}.

At redshifts $z\lesssim4$ the QLF is generally found to be well represented by a double power law \citep{Boyle1988},
\begin{equation}
 \Psi(M,z) = \frac{\Psi^{\star}}{10^{0.4(\alpha+1)(M-M^{\star})}+10^{0.4(\beta+1)(M-M^{\star})}} \ ,
\end{equation}
where $\Psi^{\star}$ is the overall normalization, $M^{\star}$ is the break magnitude between the power laws, $\alpha$ is the faint end, and $\beta$ is the bright-end slope. All parameters could possibly evolve with redshift and a variety of evolutionary models has been proposed.

However, the ELQS quasar sample probes only the bright end of the QLF. 
As discussed before, the break magnitude at redshifts probed by the ELQS \citep[see][their Figure\,19]{McGreer2013} will evolve strongly from $M^{\star}_{1450}\approx-25.6$ at $z=2.8$ to $M^{\star}_{1450}\approx-26.5$ at $z=4.5$. In all cases the break magnitude is above the faint luminosity limit of our survey $M_{1450}<-27$. 

Therefore, we assume a fixed break magnitude of $M^{\star}_{1450}=-26$ and parameterize the QLF using a single power law, probing only the bright-end slope,
\begin{equation}
 \Psi(M,z) = \Psi^{\star}(z) \cdot 10^{-0.4(\beta+1)(M-M_{1450}^{\star})} \ .
\end{equation}
We further allow for redshift evolution of the normalization $\Psi^{\star}(z)$ following an exponential,
\begin{equation}
 \log[\Psi^{\star}(z)] = \log[\Psi^{\star}_0] + \gamma \cdot z \ ,
\end{equation}
where $\Psi^{\star}_0$ is the normalization at $z=0$ and $\gamma$ is a parameter of the exponential function. The independent redshift and magnitude evolution of this QLF parameterization is supported by our analysis in Section\,\ref{sec_corrtest}. Furthermore, the evolution in $\Psi^{\star}(z)$ is analogous to the redshift evolution of the spatial density of the QLF $\rho(z)$ in Section\,\ref{sec_cminus}.

We calculate the maximum likelihood estimates using the capabilities of \texttt{simqso} \citep{McGreer2013}. 
While we do impose the 2MASS PSC magnitude and error limits to our sample, we first do not restrict the absolute magnitude further and choose integration boundaries of $M_{1450}=-31.5$ to $-27$. For this case we calculate the fit over the full redshift range ($2.8\leq z \leq4.5$) and then restrict the sample to higher redshifts ($3.0\leq z \leq 4.5$).
Then we impose an absolute magnitude limit of $M_{1450}=-28$ on our quasar sample and change the integration boundaries for the fit accordingly ($-31.5 \leq M_{1450} \leq -28$). We again calculate this fit for an both redshift ranges. The best-fit parameters for all fits are shown in Table\,\ref{tab_mlqlffit}. For all fit parameters we quote the $1\sigma$ uncertainties. In addition, we present the $3\sigma$ uncertainties for the bright-end slope $\beta$ in parenthesis. We also display the maximum likelihood fit for the whole redshift and luminosity range, the first row of Table\,\ref{tab_mlqlffit}, as the solid red lines in Figure\,\ref{fig_binnedqlf} in comparison to the binned QLF.

While the uncertainties on the fit parameters increase strongly toward the more restricted samples, the best-fit values are all consistent within their $1\sigma$ uncertainties. Furthermore the slope of the power law and the evolution of the normalization are also in agreement with our fit to the results from the $\text{C}^{-}$ estimator at the $1\sigma$ level .

However, there are slight differences in the fit results within the $1\sigma$ uncertainties. The fits over the full redshift range are very similar in all three parameters. This does not hold true for the higher redshift subsample ($z=3.0-4.5$), where the parameters between the sample over the full range, $M_{1450}=31.5$ to $-27$, and the brighter subsample, $M_{1450}=31.5$ to $-28$, result in notable differences in the normalization ,$\Delta \log[\Psi_0^{\star}] = 0.41$, and in $\gamma$, $\Delta \gamma =0.1$.

Overall, the maximum likelihood analysis encourages a steep bright-end slope with best-fit values around $\beta\approx-4.0$ and excludes slopes flatter than $\beta=-2.94$ at the $3\sigma$ level.
This result contrasts some earlier work by \citet{Koo1988, Schmidt1995, Fan2001b, Richards2006, Masters2012}, who found bright-end slopes that are around $\beta\approx-3$ at $z=2-3$ and flatten toward higher redshifts ($\beta\approx-2.5$ at $z=5$).
However, there is a range of studies at lower \citep{Croom2009, Ross2013} and higher redshifts \citep{Jiang2008, Willott2010a, McGreer2013, Yang2016}, which found the bright-end slope to be quite steep with values around $\beta\approx-3.5$.

The evolution of the overall normalization (density evolution) of the QLF shows a decline at the bright end with best-fit values of $\gamma\approx-0.37$ for our choice of parameterization. A range of studies on the QLF show that the normalization of the QLF declines steeply toward higher redshifts \citep{Fan2001b, Jiang2008, McGreer2013, Ross2013, Yang2016}. 
Our best-fit result of $\gamma$ argues for a more moderate decline of the normalization at the brightest end compared to \citet[][$\gamma\approx-0.5$ at $z=3-5$]{Fan2001b}, \citet[][$\gamma\approx-0.8$ at $z=4.7-5.4$]{Yang2016}, and \citet[][$\gamma\approx-0.7$ at $z=2.2-3.5$]{Ross2013}. However, the uncertainties are still large and do agree with $\gamma=-0.5$ at the $1\sigma$ level. 

While our quasar sample is limited to the bright end of the distribution ($M_{1450}\leq-27$), it should be noted that none of the previous estimations of the QLF extended to luminosities as bright as we sample here. Therefore, it is possible that differences between our best-fit values and other estimates are a result of a luminosity dependence of the QLF model.

\section{Conclusion}\label{sec_conclusion}

In a first publication \citep{Schindler2017}, we discussed the incompleteness of extremely bright quasars in the SDSS and BOSS spectroscopic surveys and presented the design of the ELQS to find previously missed quasars. In this paper, we present the results of the ELQS survey in the SDSS North Galactic Cap footprint (ELQS-N).
Spectroscopic followup of ELQS-N quasar candidates has been completed using the VATT and MMT telescopes and we have presented the ELQS-N quasar sample in this work. 

A total of 340 good ELQS-N candidates were selected with our quasar selection method, of which 252 were known quasars in the literature (21 at $z<2.8$, 231 at $z\geq2.8$). We followed up 88 candidates with optical spectroscopy and discovered 39 new quasars in our targeted redshift range. In summary, the final ELQS-N quasar sample includes 270 quasars with $m_{i}\leq18.0$ at $2.8\leq z \leq4.5$. The 39 new quasars are presented in Table\,\ref{tab_elqs_spring_newqsos}.

We have cross matched the full ELQS-N catalog with FIRST, \textit{GALEX} GR6/7, and \textit{ROSAT} 2RXS and XMMSL2 catalogs pre-matched to AllWISE. While we could not identify cross matches to XMMSL2 in this sample, we find 8 matches to the prematched AllWISE \textit{ROSAT} 2RXS catalog, 34 matches to FIRST and 38 counterparts in \textit{GALEX}. If we count all quasars with radio detections as radio-loud, then we can estimate that the ELQS-N sample has a radio-loud fraction of $\approx12.6\%$. This value is in rough agreement with predictions of \cite{Jiang2007}.
Out of the newly discovered 39 quasars, 8 have valid \textit{GALEX} near-UV magnitudes and 2 were detected in the far-UV as well, providing promising new candidates to investigate the He-reionization of the universe.

The new ELQS-N quasars include six objects that show BAL features. Using the qualitative information on BALs from the literature and the DR12 quasar catalog, we could identify 57 out of 262 ELQS-N quasars to be BALs, resulting in a BAL fraction of $\sim22\%$. Eight objects did not have sufficient information to allow for a visual classification. 
Our estimated BAL fraction is higher compared to results on other samples uncorrected for selection effects \citep{Trump2006, Maddox2008, Allen2011}. It is unclear whether this is a result of our infrared-based quasar selection \citep{Dai2008, Maddox2008}, a suggested redshift evolution of the BAL fraction \citep{Allen2011}, a luminosity dependence of the BAL fraction, or any combination of these effects.

We carefully analyze the completeness of our quasar selection using a grid of simulated quasars. While the \textit{JKW2} color cut proves to be highly inclusive, independent of redshift or apparent magnitude, the photometric limits on the 2MASS survey, our random forest classification, and the random forest redshift estimation limit our survey completeness. However, between redshifts $3.0\lesssim z \lesssim 5.0$ and at $m_{i}\lesssim17.5$ the ELQS completeness is generally above $\sim 70\%$. 

Using the estimated completeness, we calculate the binned QLF in four redshift bins from $2.8\leq z \leq4.5$ (Figure\,\ref{fig_binnedqlf}) and show that we can extend previous estimates by one magnitude toward the bright end. Our binned QLF shows a steeper slope, when compared to the SDSS DR3 QLF fit of \citet{Richards2006} in all four redshift bins.

We further analyze the marginal distributions of the QLF as a function of magnitude and redshift (Figure\,\ref{fig_marg_dist}) and estimate a bright-end slope of $\alpha=-3.98\pm0.18$ for quasars with $M_{1450}\leq-27.7$ over $z=2.8-4.5$. 
A maximum likelihood fit to the full quasar sample at $z=2.8-4.5$ using a single power-law QLF with redshift evolution in the normalization returns a steep bright-end slope of $-3.96^{+0.21}_{-0.22}$. Furthermore the maximum likelihood analysis encourages slopes steeper than $\beta=-2.94$ at the $3\sigma$ level.

A range of studies \citep{Koo1988, Schmidt1995, Fan2001b, Richards2006} have suggested a flattening of the bright-end slope at $z\approx3-5$ with values around $\beta\approx-2.5$. However, more recent work at $z=0.4-2.6$ \citep{Croom2009, Ross2013} and at redshift $z\geq5$ \citep{Jiang2008, Willott2010a, McGreer2013, Yang2016} found that the bright-end slope remains fairly steep $\beta\approx-3.5$. 
With our results, probing the intermediate redshift range $z=2.8-4.5$, it now seems likely that the bright-end slope of the QLF remains steep from $z=0.4$ up to $z\geq5$.

Our constraints on the normalization (density evolution) of the QLF show a more moderate decline with $\gamma\approx-0.3$ toward higher redshifts as previous works suggest \citep{Fan2001b, Ross2013, Yang2016}. Unfortunately, the uncertainties on our best-fit parameters are too large to allow for further investigation.

In a forthcoming and concluding publication we will summarize the results of the ELQS over the entire SDSS footprint, provide a full analysis of the QLF, and discuss further implications for the bright quasar population.

\subsection*{Acknowledgements}
The authors thank the staffs of the MMT and VATT telescopes for enabling many of the ELQS observations.
J.-T.S. and X.F. acknowledge support from the U.S. NSF grant AST 15-15115 and NASA ADAP grant NNX17AF28G. 

This publication makes use of data products from the Two Micron All Sky Survey, which is a joint project of the University of Massachusetts and the Infrared Processing and Analysis Center/California Institute of Technology, funded by the National Aeronautics and Space Administration and the National Science Foundation.
This publication makes use of data products from the Wide-field Infrared Survey Explorer, which is a joint project of the University of California, Los Angeles, and the Jet Propulsion Laboratory/California Institute of Technology, funded by the National Aeronautics and Space Administration.

Funding for the Sloan Digital Sky Survey IV has been provided by the Alfred P. Sloan Foundation, the U.S. Department of Energy Office of Science, and the Participating Institutions. SDSS acknowledges support and resources from the Center for High-Performance Computing at the University of Utah. The SDSS website is \url{www.sdss.org}.

SDSS is managed by the Astrophysical Research Consortium for the Participating Institutions of the SDSS Collaboration, including the Brazilian Participation Group, the Carnegie Institution for Science, Carnegie Mellon University, the Chilean Participation Group, the French Participation Group, Harvard-Smithsonian Center for Astrophysics, Instituto de Astrofísica de Canarias, The Johns Hopkins University, Kavli Institute for the Physics and Mathematics of the Universe (IPMU) / University of Tokyo, Lawrence Berkeley National Laboratory, Leibniz Institut f\"ur Astrophysik Potsdam (AIP), Max-Planck-Institut f\"ur Astronomie (MPIA Heidelberg), Max-Planck-Institut f\"ur Astrophysik (MPA Garching), Max-Planck-Institut f\"ur Extraterrestrische Physik (MPE), National Astronomical Observatories of China, New Mexico State University, New York University, University of Notre Dame, Observatório Nacional / MCTI, The Ohio State University, Pennsylvania State University, Shanghai Astronomical Observatory, United Kingdom Participation Group, Universidad Nacional Autónoma de México, University of Arizona, University of Colorado Boulder, University of Oxford, University of Portsmouth, University of Utah, University of Virginia, University of Washington, University of Wisconsin, Vanderbilt University, and Yale University.

This research has made use of the NSED, which is operated by the Jet Propulsion Laboratory, California Institute of Technology, under contract with the National Aeronautics and Space Administration.

This research made use of Astropy, a community-developed core Python package for Astronomy (Astropy Collaboration, 2013, \url{http://www.astropy.org}). In addition, Python routines from scikit-learn\citep{scikit-learn}, SciPy \citep{scipy}, Matplotlib \citep{matplotlib}, and Pandas \citep{pandas} were used in the quasar selection, data analysis and creation of the figures.

\appendix

\section{The ELQS-N Quasar Catalog} \label{app_qso_catalog}
The ELQS-N quasar catalog is available as a machine-readable table online. It has 57 columns, detailed in Table\,\ref{tab_elqs_cat_cols}.

\begin{table*}[htp]
 \centering 
 \caption{Description of the ELQS-N quasar catalog table}
 \label{tab_elqs_cat_cols}
 \begin{tabular}{cccp{8cm}}
  \tableline
  Column & Column Name & Unit & Description \\
  \tableline
  1 & WISE &  - & Designation of the \textit{WISE} AllWISE survey\\
  2 & RAdeg-SDSS & deg & SDSS DR13 Right Ascension, decimal degrees (J2000) \\
  3 & DEdeg-SDSS & deg & SDSS DR13 Declination, decimal degrees (J2000) \\
  4 & RAh & h & Hour of SDSS DR13 Right Ascension (J2000)  \\
  5 & RAm & min & Minute of SDSS DR13 Right Ascension (J2000)  \\
  6 & RAs & s & Second of SDSS DR13 Right Ascension (J2000)  \\
  7 & DE- & - & Sign of SDSS DR13 Declination (J2000)  \\
  8 & DEd & deg & Degree of SDSS DR13 Declination (J2000)   \\
  9 & DEm & arcmin &  Arcminute of SDSS DR13 Declination (J2000)   \\
  10 & DEs & arcsec & Arcsecond of SDSS DR13 Declination (J2000)   \\
  11 & RAdeg-WISE & deg & \textit{WISE} Right Ascension, decimal degrees (J2000) \\
  12 & DEdeg-WISE & deg &  \textit{WISE} Declination, decimal degrees (J2000) \\
  13 & Ref & - & Reference to the quasar classification \\
  14 & z-Ref & - & Best redshift of the quasar according to the reference \\
  15 & M1450 & mag & Absolute magnitude, $1450\text{\AA}$, calculated using the k-correction determined in this work\\
  16 &  sel-prob & - & Selection probability according to our completeness calculation \\
  17-26 & [band]mag & mag & Dereddened AB magnitudes of the SDSS DR13 ugriz, 2MASS JH$\rm{K}_{\rm{s}}$ and AllWISE W1, W2 bands (bands = [u,g,r,i,z,J,H,K,W1,W2]). It should be noted that all SDSS magnitudes are PSF magnitudes in the SDSS Asinh magnitude system.\\
  27-36 & e\_[band]mag & mag & $1\sigma$ errors on the AB magnitudes. \\
  37 &E(B-V) & mag & E(B-V) color excess \\
  38 & Ai & mag & Extinction in the SDSS i-band\\ 
  39 & FIRST & - & Boolean to indicate successful matches with the FIRST catalog \\
  40 & Sep-FIRST & arcsec & Distance of the FIRST source relative to the SDSS position \\
  41 & Flux-FIRST-peak & mJy/bm & FIRST peak flux, mJy/beam\\
  42 & Flux-FIRST-int & mJy & FIRST integrated flux, mJy\\
  43 & e\_Flux-FIRST-RMS & mJy/bm & RMS error on the FIRST flux, mJy/beam\\
  44 & GALEX & - &  Boolean to indicate successful matches with the \textit{GALEX} GR6/7 catalog\\
  45 & Sep-GALEX & arcsec & Distance of the \textit{GALEX} GR6/7 match relative to the SDSS position \\
  46 & NUVmag & mag & \textit{GALEX} GR6/7 Near-UV flux, magnitudes \\
  47 & e\_NUVmag & mag & Uncertainty in NUVmag \\
  48 & FUVmag & mag & \textit{GALEX} GR6/7 Far-UV flux, magnitudes \\
  49 & e\_FUVmag & mag & Uncertainty in FUVmag \\
  50 & TRXS & - & Boolean to indicate successful matches to the \textit{ROSAT} 2RXS AllWISE counterparts \\
  51 & Sep-TRXS & arcsec & Match distance between the ELQS AllWISE position to the \textit{ROSAT} 2RXS AllWISE position. The distance values are often 0 or otherwise extremely small, because the positions match to numerical accuracy.\\
  52 & f\_TRXS & - & A flag indicating the most probable AllWISE \textit{ROSAT} 2RXS cross-match with 1. This is the case for all matched objects.\\
  53 & Flux-TRXS & $\rm{erg}\,\rm{cm}^{-2}\,\rm{s}^{-1}$ &  2RXS flux \\
  54 & e\_Flux-TRXS &  $\rm{erg}\,\rm{cm}^{-2}\,\rm{s}^{-1}$ & Uncertainty in Flux-TRXS\\
  56 & BAL & 1/0/-1 & Broad absorption line flag, indicating whether the object is visually identified as a broad absorption line (BAL) quasar ($1=$ BAL quasars, $0=$ quasars, $-1=$ no visual classification).\\
  57 & QLF & - & Boolean to indicate whether the quasar is included in the estimation of the quasar luminosity function (Section\,\ref{sec_lumfun}).\\
  \tableline
 \end{tabular}
\end{table*}

\section{Flux Ratios of Simulated Quasars}\label{app_flux_ratios}
\begin{figure*}[h]
 \includegraphics[width=\textwidth]{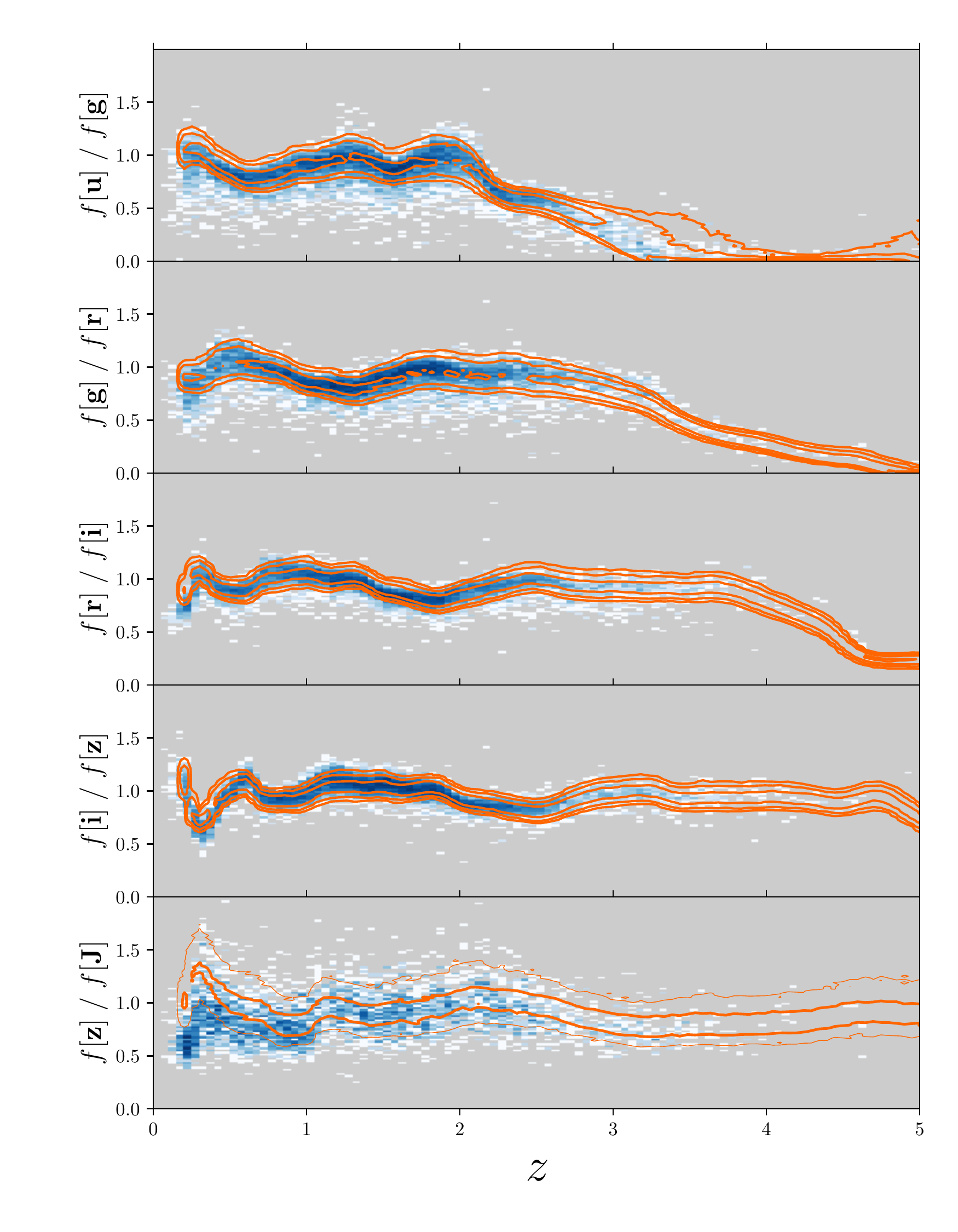}
 \caption{This figure shows the flux-ratio redshift distributions of the known quasars in the DR7Q and DR14Q with $m_{i}\leq18.5$ as the white to blue density map. The over-plotted orange contours correspond to the sample of simulated quasars (see Sec.\,\ref{sec_simqso}). We show the $ug, gr, ri, iz,$ and $zJ$ flux ratios of the SDSS and 2MASS bands.}
 \label{fig_frz_a}
\end{figure*}

\begin{figure*}[h]
 \includegraphics[width=\textwidth]{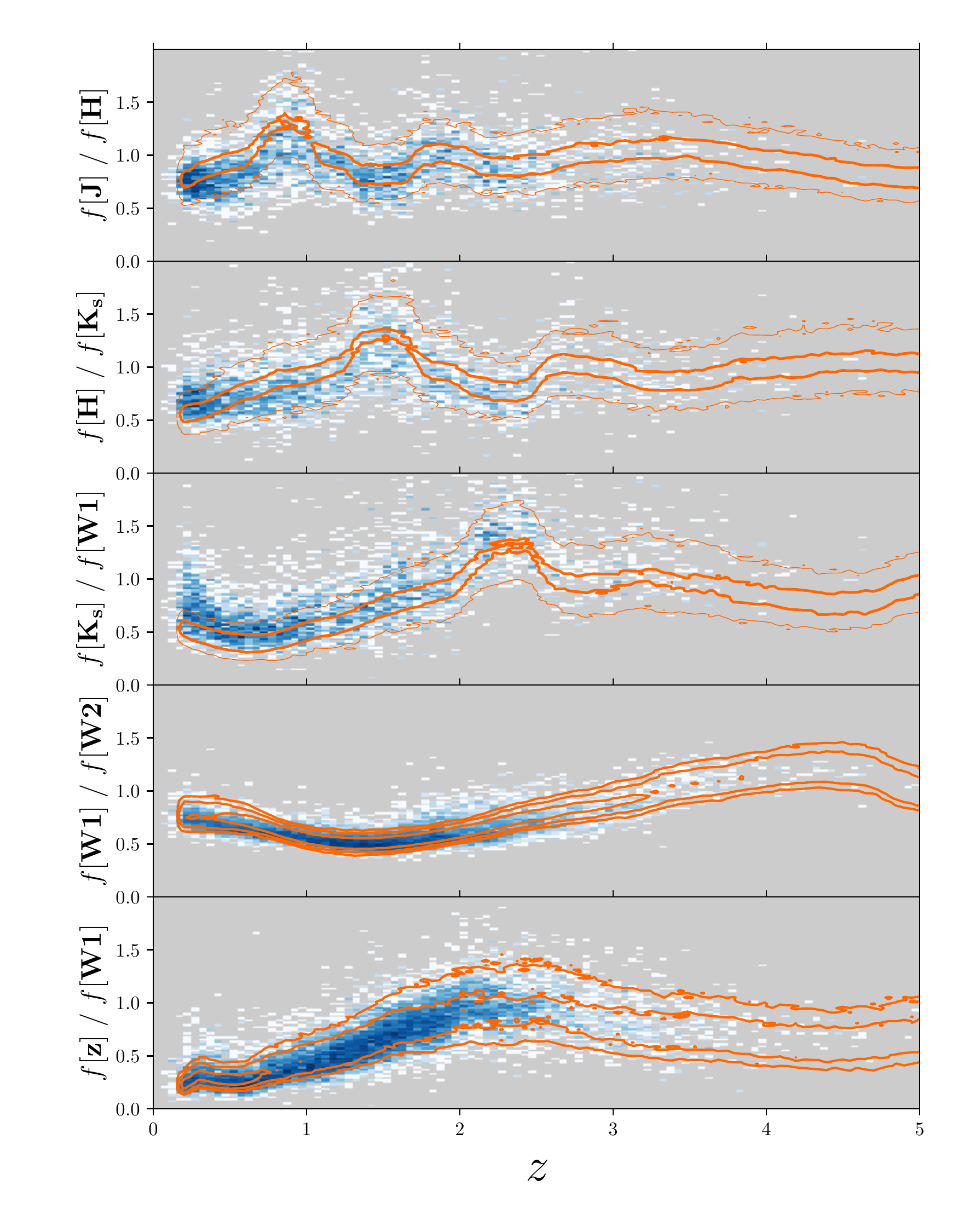}
 \caption{This figure shows the flux-ratio redshift distributions of the known quasars in the DR7Q and DR14Q with $m_{i}\leq18.5$ as the white to blue density map. The over-plotted orange contours correspond to the sample of simulated quasars (see Sec.\,\ref{sec_simqso}). We show the $JH,HK_{\rm{s}}, K_{\rm{s}}W1, W1W2, $ and $zW1$ flux ratios of the SDSS, 2MASS and \textit{WISE} bands.}
 \label{fig_frz_b}
\end{figure*}

We present two figures (Figures\,\ref{fig_frz_a} and \ref{fig_frz_b}) showing the $ug, gr, ri, iz, zj, JH, HK_{\rm{s}}, K_{\rm{s}}W1, W1W2$, and $zW1$ (SDSS, AllWISE and 2MASS) flux ratios as a function of redshift for an empirical quasar sample from the DR7Q/DR12Q quasar catalogs and the simulated quasar set used to calculate the completeness (see Section\,\ref{sec_completeness}). The DR7Q/DR12Q quasars are displayed as a white to blue density map, while we show the simulated quasars in orange density contours. For all quasars we use a faint magnitude limit on both empirical and simulated quasars of $m_{i}<18.5$.

The figures demonstrate that flux ratios of the simulated quasars qualitatively follow the ones from the empirical quasars as a function of redshift. Only at $z<1$ do the $zJ$ and $K_{\rm{s}}W1$ flux ratios of the simulated quasars deviate from the empirical data set. 

\bibliographystyle{apj}

\end{document}